\RequirePackage[dvipsnames]{xcolor}
\RequirePackage{tikz}
\usetikzlibrary{positioning}

\documentclass[sn-mathphys]{sn-jnl}% Math and Physical Sciences Reference Style
\usepackage{amsmath}
\usepackage{mathtools}
\usepackage{enumitem}
\usepackage{graphicx}
\usepackage{siunitx}
\usepackage{float}
\usepackage{placeins}
\jyear{2023}%

\theoremstyle{thmstyleone}%
\theoremstyle{thmstyletwo}%

\theoremstyle{thmstylethree}%

\begin{document}

\title[Geometry of charged rotating discs of dust in Einstein-Maxwell theory]{Geometry of charged rotating discs of dust in Einstein-Maxwell theory}


\author*[1]{\fnm{David} \sur{Rumler}}\email{david.rumler@uni-jena.de}

\author[1]{\fnm{Andreas} \sur{Kleinwächter}}%\email{andreas.kleinwaechter@uni-jena.de}

\author[1]{\fnm{Reinhard} \sur{Meinel}}%\email{reinhard.meinel@uni-jena.de}

\affil*[1]{\orgdiv{Theoretisch-Physikalisches Institut}, \orgname{Friedrich-Schiller-Universität Jena}, \orgaddress{\street{Max-Wien-Platz 1}, \mbox{\postcode{07743} \city{Jena}}, \country{Germany}}}


\abstract{Within the framework of Einstein-Maxwell theory geometric properties of charged rotating discs of dust, using a post-Newtonian expansion up to tenth order, are discussed. Investigating the disc's proper radius and the proper circumference allows us to address questions related to the Ehrenfest paradox. In the Newtonian limit there is an agreement with a rotating disc from special relativity. The charged rotating disc of dust also possesses material-like properties. A fundamental geometric property of the disc is its Gaussian curvature. 
%Taking different limits, it coincides with the Gaussian curvature of the limiting cases for which an exact analytic solution is available. 
%The Gaussian curvature of limiting cases for which an exact analytic solution is available coincides with the Gaussian curvature in the corresponding limits.
%We compute the Gaussian curvature of the charged disc of dust as well as the analytic limiting cases (charged) Maclaurin disc, electrically counterpoised dust disc and the uncharged disc of dust and verify their agreement in appropriate limits. The exact analytic solutions of those limiting cases coincide with the corresponding limits of the Gaussian curvature of the charged disc of dust.
%By additionally computing the Gaussian curvature of  the analytic limiting cases (charged rotating) Maclaurin disc, electrically counterpoised dust disc and uncharged rotating disc of dust, the result for Gaussian curvature of the charged rotating disc of dust is verified in those limits.
The result obtained for the charged rotating disc of dust is checked by additionally calculating the Gaussian curvature of the analytic limiting cases (charged rotating) Maclaurin disc, electrically counterpoised dust-disc and uncharged rotating disc of dust.
We find that by increasing the disc's specific charge there occurs a transition from negative to positive curvature.}




\maketitle

%\tableofcontents

\section{Introduction}\label{sec1}

In 1909 Ehrenfest formulated a famous paradox concerning a rigidly rotating disc (or a cylinder in the original version) within special relativity \cite{Ehrenfest}. He pointed out that special relativistic rigidity, introduced by Born \cite{Born}, easily leads to two contradicting statements about radius and circumference measured by a non-rotating observer before and after the disc is set into rotation.

The paradox gave rise to a long lasting debate on how to resolve it, see, e.g., \cite{Gron2004}. Numerous physicists came up with solution approaches, however, they turned out to be wrong in most cases. Gr\o{}n presented in 1975 a resolution solely based on kinematics \cite{Gron1975}. Initially, it was not clear whether this would be sufficient or whether a dynamical treatment would be necessary to solve the problem  completely. Today, most physicists (including the authors of this paper) accept Gr\o{}n's kinematic considerations as the correct way to resolve Ehrenfest's paradox.

Nevertheless there are related questions to the paradox, as how the ratio of circumference to radius is observed from the point of view of non-rotating versus co-rotating observers (on the disc), how those measurements change when the rotation speed is altered and how material properties affect more realistic discs. 

With the help of the available semi-analytic solution of the charged rotating disc of dust, which is a concrete, physically relevant solution of the Einstein-Maxwell equations in terms of a post-Newtonian expansion \cite{Breithaupt_2015}, we can tackle the above questions in a very direct way.

Motivated by those geometric investigations it is also natural to ask about the intrinsic curvature of the charged rotating disc of dust. For a two-dimensional disc the intrinsic curvature is simply the Gaussian curvature. In particular, the post-Newtonian expansion allows us not only to study special relativistic effects, but also investigate the influence of higher order corrections coming from general relativity. A key observation is that the specific charge of the disc acts like a regulator to adjust the curvature of the disc. More precisely, by increasing the specific charge from $0$ to $1$ there is a critical value at which a transitions from negative to positive curvature occurs.

The present paper is structured as follows. In section \ref{sec2} we briefly introduce the mathematical formulation of the disc problem in terms of a boundary value problem, followed by a short overview of the used post-Newtonian expansion in section \ref{sec3}. After defining the proper radius and the proper circumference of the rotating disc in section \ref{sec4}, we discuss the above mentioned questions related to the Ehrenfest paradox based on the charged rotating disc of dust. \mbox{Section \ref{sec5}} is devoted to the curvature of the disc. The result will be compared with the ones for a Maclaurin disc within Newtonian theory, a disc with a concrete electrically counterpoised dust-configuration (having maximal charge) and the uncharged disc of dust.

\section{Boundary value problem for the charged disc of dust}\label{sec2}

%We consider an infinitesimally thin disc of dust (perfect fluid with vanishing pressure) in thermodynamic equilibrium with specific charge $\epsilon = \frac{\rho_{\text{el}}}{\mu} \in [-1,1]$ and constant angular velocity $\Omega$ \cite{Palenta_2013, Breithaupt_2015, Meinel_2015}. 
We consider an equilibrium configuration of an infinitesimally thin disc of dust (perfect fluid with vanishing pressure) with constant specific charge \mbox{$\epsilon = \frac{\rho_{\text{el}}}{\mu} \in [-1,1]$ and constant angular velocity $\Omega$ \cite{Palenta_2013, Breithaupt_2015, Meinel_2015}. 
$\rho_{\text{el}}$} and $\mu$ denote the charge density and the baryonic mass density of the disc, respectively. 

The corresponding disc problem is formulated in the framework of Einstein-Maxwell equations\footnote{We use units in which $c=G=4\pi\epsilon_{0}=1$.}:
\begin{gather}
	R_{ab}-\frac{1}{2}R\,g_{ab} = 8\pi T_{ab} \,,\\
	%F_{[ab{;}c]}=0 \,, 
	\quad {F^{ab}}_{;\negthickspace b} = 4\pi j^a \,,
\end{gather}
where
\begin{align}
	&T_{ab} = T_{ab}^{(\text{em})} + T_{ab}^{(\text{dust})} = \frac{1}{4\pi}\left( F_{ac}{F_b}^c - \frac{1}{4}g_{ab}F_{cd}F^{cd} \right) + \mu u_au_b \,,\\
	&j^a = \rho_{\text{el}}u^a = \epsilon\mu u^a \,, \\
	&F_{ab} = A_{b;\negthickspace a} - A_{a;\negthickspace b} = A_{b,a} - A_{a,b} \,.
\end{align}
To the energy momentum tensor, $T_{ab}$, both the electromagnetic field and matter (dust) contribute.
$j^a$ represents a purely convective four-current density, where $u^a$ is the four-velocity. As is common practice, the field strength tensor $F_{ab}$ is expressed in terms of the four-potential $A_a$.

%Due to the thermodynamic equilibrium of the charged dust, spacetime is stationary, implying the existence of an associated Killing vector $\boldsymbol{\xi}$. We also demand axisymmetry, characterized by another Killing vector $\boldsymbol{\eta}$ that commutes with $\boldsymbol{\xi}$. Furthermore, the problem is reflection symmetric with respect to the equatorial plane of the disc.
For an equilibrium configuration, spacetime is stationary, implying the existence of an associated Killing vector $\boldsymbol{\xi}$. We also demand axisymmetry, characterized by another Killing vector $\boldsymbol{\eta}$ that commutes with $\boldsymbol{\xi}$. Furthermore, the problem is reflection symmetric with respect to the equatorial plane of the disc.

Exploiting stationarity and axisymmetry the line element can globally be written in terms of Weyl-Lewis-Papapetrou coordinates:
\begin{equation}\label{eq:bvp.4}
	\mathrm{d}s^2 = f^{-1}\left[ h\left( \mathrm{d}\rho^2 + \mathrm{d}\zeta^2 \right) + \rho^2\mathrm{d}\varphi^2 \right] - f\left( \mathrm{d}t + a\, \mathrm{d}\varphi \right)^2 \,.
\end{equation}
Using Lorenz-gauge the four-potential $A_a$ takes the form:
\begin{equation}
	A_a = (0,0,A_{\varphi}, A_t)\,.% \quad \text{with} \quad F_{ab} = A_{b;a} - A_{a;b} = A_{b,a} - A_{a,b}. %\footnote{With the above definition of the field strength tensor, $F_{ab} = A_{b;a} - A_{a;b}$, the homogeneous Maxwell equation is fulfilled automatically.}
\end{equation} 
The coordinates $t$ and $\varphi$ used in the line element are adapted to the Killing vectors, such that
\begin{equation}
	\boldsymbol{\xi}=\frac{\partial}{\partial t} \quad \text{and} \quad \boldsymbol{\eta}=\frac{\partial}{\partial \varphi} \,.
\end{equation}
As a consequence, the metric functions $f$, $h$, and $a$ as well as the potentials $A_{\varphi}$ and $A_t$ depend on the coordinates $\rho$ and $\zeta$ only.
Instead of $f$ and $h$ we will also make use of the functions $U$ and $k$, where $f=e^{2U}$ and $h=e^{2k}$.

In the subsequent discussion we heavily rely on the co-rotating frame of reference defined by $\varphi'=\varphi-\Omega t$. As usual, the covariance of the metric and the four-potential imply the following transformation laws:
\begin{align}\label{eq:bvp.5}
	&f'=f\left[\left(1+\Omega a\right)^2 - \Omega^2\rho^2 f^{-2}\right] \,, \quad \left(1-\Omega a'\right)f'=\left(1+\Omega a\right)f \,, \quad \frac{h'}{f'}=\frac{h}{f} \,, \\
	&A'_{\varphi'}=A_{\varphi} \,, \quad A'_{t}=\Omega A_{\varphi}+A_t \,.
\end{align}

It should be noted that the mass density $\mu$ is related to the surface mass density $\sigma$ via
\begin{equation}
	\mu = \frac{f}{h}\sigma(\rho)\delta(\zeta) \,.
\end{equation}
However, $\sigma$ is not coordinate independent in contrast to the proper surface mass density
\begin{equation}
	\sigma_\text{p} = \sqrt{\frac{f}{h}}\sigma \,.
\end{equation}

In (electro-)vacuum, assuming axisymmetry and stationarity, the combined Einstein-Maxwell equations in terms of $f$, $h$, $a$, $A_{\varphi}$ and $A_t$, that can be found in \cite{Palenta_2013}, can be reduced to the Ernst equations \cite{PhysRev.168.1415}:
\begin{align}
	\label{eq:bvp.1}
	\left( \Re \mathcal{E} + \lvert \Phi \rvert^2 \right) \Delta \mathcal{E} &= \left( \nabla\mathcal{E} + 2\bar{\Phi}\nabla\Phi \right) \cdot \nabla\mathcal{E} \,, \\
	\label{eq:bvp.2}
	\left( \Re \mathcal{E} + \lvert \Phi \rvert^2 \right) \Delta \Phi &= \left( \nabla\mathcal{E} + 2\bar{\Phi}\nabla\Phi \right) \cdot \nabla\Phi \,,
\end{align}
where 
\begin{equation}
	\Phi = \alpha + \mathrm{i}\beta \quad \text{and} \quad \mathcal{E} = \left(f-\lvert\Phi\rvert^2\right) + \mathrm{i}b \,,
\end{equation}
with $\alpha\coloneqq \Re\Phi=-A_t$, are the Ernst potentials.
The potentials $\beta$ and $b$ are defined via
\begin{equation}\label{eq:bvp.6a}
	\beta_{,\rho} = \frac{f}{\rho}\left( A_{\varphi,\zeta} - aA_{t,\zeta} \right) \,, \quad \beta_{,\zeta} = -\frac{f}{\rho}\left( A_{\varphi,\rho} - aA_{t,\rho} \right)
\end{equation}
and
\begin{equation}\label{eq:bvp.6b}
	b_{,\rho} = - \frac{f^2}{\rho}a_{,\zeta} - 2\left( \beta A_{t,\rho}-A_t\beta_{,\rho} \right)        \,, \quad b_{,\zeta} = \frac{f^2}{\rho}a_{,\rho} - 2\left( \beta A_{t,\zeta}-A_t\beta_{,\zeta} \right) \,.
\end{equation}
$\nabla$ and $\Delta$ are meant as operators in 3-dimensional Euclidean space with cylindrical coordinates $\rho,\zeta$ and $\varphi$.
Note that the integrability conditions for (\ref{eq:bvp.6a}) and (\ref{eq:bvp.6b}) are automatically satisfied as a consequence of the Einstein-Maxwell equations. Vice versa, (\ref{eq:bvp.6a}) and (\ref{eq:bvp.6b}) may be used to obtain $a$ and $A_{\varphi}$ from the solutions to the Ernst equations by a path-independent line integration.
%In the Ernst equations the function $h$ is absent, since it can be determined by a path-independent line integration from the other functions.
The function $h$ can be determined by a path-independent line integration from the other functions as well.

Using reflection symmetry the following boundary conditions on the disc, i.e.\ for \mbox{$0\leq \rho \leq \rho_0$} and $\zeta=0$, where $\rho_0$ is the coordinate radius, can be derived:
\begin{equation} \label{eq:bvp.3}
	b'=0\,, \quad \beta'=0\,, \quad \left(e^{U'}-\epsilon A'_t\right)_{,\rho}=0\,, \quad \left(A'_t-\epsilon e^{U'}\right)_{,\zeta}=0 \,. 
\end{equation}
The Ernst equations (\ref{eq:bvp.1}) and (\ref{eq:bvp.2}) together with the boundary conditions (\ref{eq:bvp.3}),  as well as the asymptotic flatness condition ($\mathcal{E}\to 1$ and $\Phi \to 0$ for \mbox{$\rho^2+\zeta^2 \to \infty$}) and the regularity condition at the rim of the disc, form a well-defined boundary value problem for the charged disc of dust.

\section{Post-Newtonian expansion}\label{sec3}

In case of the rotating disc of dust without charge the corresponding boundary value problem could be solved analytically \cite{PhysRevLett.75.3046, RFE}. For the charged rotating disc of dust a semi-analytic solution in terms of a post-Newtonian expansion up to tenth order is available \cite{Palenta_2013, Breithaupt_2015}.

This expansion uses a relativity parameter $\gamma$ \cite{bardeen}, defined by
\begin{equation}\label{eq:pne.1}
	\gamma \coloneqq 1-\sqrt{f_c} \,, \quad \text{with} \quad f_c \coloneqq f(\rho=0,\zeta=0) \,.
\end{equation}
Equivalently, $\gamma$ can be expressed in terms of the redshift, $Z_c$, of a photon emitted at $\rho=0$, $\zeta=0$ and observed at $\rho^2+\zeta^2 \to \infty$:
\begin{equation}\label{eq:pne.2}
	\gamma = \frac{Z_c}{1+Z_c} \,.
\end{equation}
From both expressions, (\ref{eq:pne.1}) and (\ref{eq:pne.2}), it is evident that $\gamma$ takes values between $0$ and $1$, where $\gamma \approx 0$ represents Minkowski spacetime and $\gamma\to 1$ corresponds to the ultra-relativistic limit. At the ultra-relativistic limit we assume the formation of black holes \cite{Breithaupt_2015}.
Note that $U$, appearing in $f_c=e^{2U_c}$, can be interpreted as a generalized Newtonian potential.

The parameter space of the charged disc solution is spanned by \mbox{$\gamma \in [0,1]$}, $\epsilon \in [0,1]$ and the coordinate radius $\rho_0$ as a scaling parameter. By restricting to positive charges we do not loose generality.\footnote{In fact, all dimensionless physical parameters describing the disc are functions of $\gamma$ and $\epsilon$, in particular $\rho_0\Omega$, $M/\rho_0$ and $J/\rho_0^2$ (with $M$ and $J$ being gravitational mass and angular momentum).}

A convenient choice of coordinates is given in terms of elliptic coordinates $\eta$ and $\nu$, defined through
\begin{equation}
	\rho = \rho_0\sqrt{\left(1-\eta^2\right)(1+\nu^2)} \,, \quad \zeta=\rho_0\eta\nu \,,
\end{equation}
with $\eta\in[-1,1]$ and $\nu\in[0,\infty]$. Additionally normalizing all dimensioned quantities with suitable powers of $\rho_0$ leads to a dimensionless formulation of the disc problem. The normalized quantities are denoted by $^*$, e.g.
\begin{equation}
	a^*=\frac{a}{\rho_0} \,, \quad A^*_{\varphi}=\frac{A_{\varphi}}{\rho_0} \quad \text{or} \quad \Omega^*=\rho_0\Omega \,.
\end{equation}
As a starting point of the post-Newtonian expansion we utilize the relation 
\begin{equation}
	\gamma = -U_N(\rho=0,\zeta=0)=\frac{{\Omega^*_N}^2}{1-\epsilon^2} \,,
\end{equation}
between the Newtonian potential $U_N$ and the angular frequency $\Omega_N$ that holds in Newtonian limit only.
As a consequence, we have \mbox{${\Omega^*}^2=\left(1-\epsilon^2\right)\gamma+\mathcal{O}(\gamma^2)$} in general (beyond the Newtonian limit). Expanding this relation implies that $\Omega^*$ is an odd function in \mbox{$g\coloneqq \sqrt{\gamma}$}, i.e.
\begin{equation}\label{eq:pne.3}
	\Omega^* = \sum_{n=0}^{\infty} \Omega^*_{2n+1}g^{2n+1} \,,
\end{equation}
where $\Omega^*_{2n+1}$ is a function of $\eta$ and $\nu$ and $\Omega^*_{1}=\sqrt{1-\epsilon^2}$.\footnote{It turns out that $\Omega$ (appropriately normalized, e.g. by $R_{0}$ introduced in section \ref{sec4}) increases monotonically with $g$ and decreases monotonically with $\epsilon$. (For both $g=0$ and $\epsilon=1$ it vanishes.)}

All the discussed potentials and metric functions can be expanded in $\Omega^*$.
%By means of symmetry arguments  it can be deduced that they behave either symmetric or antisymmetric under the transformation $\Omega^* \to -\Omega^*$ that changes the sense of rotation.
Examining their symmetry properties reveals a purely symmetric or antisymmetric behaviour under the transformation $\Omega^* \to -\Omega^*$ that changes the sense of rotation.
Therefore, they can also be written as a series in $g$, containing only even or odd powers of $g$, respectively. For the metric functions we get:
\begin{equation}\label{eq:pne.expansions}
	f = 1 + \sum_{n=1}^{\infty}f_{2n}g^{2n} \,, \quad h = 1 + \sum_{n=2}^{\infty}h_{2n}g^{2n} \,, \quad a^* = \sum_{n=1}^{\infty}a^*_{2n+1}g^{2n+1} \,.
\end{equation}
All the coefficient functions $f_{2n}$, $h_{2n}$ and $a^*_{2n+1}$ depend on the coordinates $\eta$ and $\nu$ only. %Those  are available up to tenth order in $n$. 

Palenta and Breithaupt were able to use this post-Newtonian expansion to derive analytic solutions for those coefficient functions. First, Palenta solved the boundary value problem in terms of this expansion up to eighth order \mbox{($n=8$)} and Breithaupt later on up to tenth \cite{Palenta_2013, Breithaupt_2015}.
The subsequent calculations and discussions of the present paper build on Breithaupt's semi-analytic solution of the charged rotating disc of dust. 

\section{Related questions to Ehrenfest's paradox}\label{sec4}

%... Some words about the Ehrenfest paradox motivating this section ...\linebreak

In 1909 Ehrenfest pointed out that rigid rotation of discs (or cylinders) within special relativity involving a period of angular acceleration inevitably leads to a paradox \cite{Ehrenfest}. 

The concept of rigidity in special relativity is defined by Born as follows:
	Each infinitesimal neighbourhood of an arbitrarily moving body, as measured by co-moving observers at each point of the body, should appear permanently undeformed \cite{Born}. 
	
Relying on this definition, Ehrenfest formulated a paradox:
	A relativistic cylinder with radius $R$ is given a rotating motion about its axis, which finally becomes constant - while satisfying Born's rigidity at all times. \linebreak Let $R''$ be the radius of the constantly rotating cylinder measured by an observer at rest, then two contradicting requirements need to be fulfilled:
	\begin{enumerate}[label=(\roman*)]
		\item $2\pi R'' < 2\pi R$ ,
		\item  $R''=R$ .
	\end{enumerate}

While to this day there is still no common agreement on its solution, there is a kinematic resolution of the paradox by Gr\o{}n most physicists do indeed accept \cite{Gron1975, Gron1977, Gron2004}. He showed that it is not possible to synchronize clocks of successive inertial rest frames momentarily at rest to points around the periphery of the rotating disc. This means that setting a disc in rotational motion, while satisfying Born's rigidity condition, is kinematically impossible.

Despite its resolution, there are still interesting questions related to Ehrenfest's paradox:
\begin{enumerate}[label=\arabic*)]
	\item How does the spatial geometry of an accelerated disc which is initially at rest evolves as measured by non-rotating  and co-rotating observers? 
	\item What is the spatial geometry of a disc that is already set into rotation as seen by non-rotating  and co-rotating observers? 
	\item A real disc has material properties, how do they influence the rotating disc?
\end{enumerate} 

The uniformly rotating disc of dust is relativistically rigid in the sense that it rotates rigidly, meaning $\Omega=\text{const.}$ \cite{Rigidity}.

In the present section we want to describe geometric quantities like proper radius and proper circumference of the rotating disc and address the above posed questions. To this purpose, we introduce the line element
\begin{equation}
	\mathrm{d}\sigma^2=h_{ik}\mathrm{d}x^i\mathrm{d}x^k
\end{equation}
with the projection tensor
\begin{equation}
	h_{ik}=g_{ik}+u_iu_k\,.
\end{equation}
This means
\begin{equation}
	\mathrm{d}s^2=\mathrm{d}\sigma^2-u_iu_k\mathrm{d}x^i\mathrm{d}x^k\,.
\end{equation}
Note that $\mathrm{d}s^2=\mathrm{d}\sigma^2$ for $\mathrm{d}x^i$ orthogonal to $u^i$, and $\mathrm{d}s^2=-\mathrm{d}\tau^2$ for $\mathrm{d}x^i=u^i\mathrm{d}\tau$ (with $\tau$ being the proper time). $\mathrm{d}\sigma^2$ can be used to measure infinitesimal proper distances on the disc. In the co-rotating frame where the four-velocity has only a fourth component (in the vicinity of the disc we consider a family of observers with the same property), we obtain
\begin{equation}\label{eq:LL}
	\mathrm{d}\sigma^2=h'_{\alpha\beta}\,\mathrm{d}x'^\alpha\mathrm{d}x'^\beta=\left(g'_{\alpha\beta}-\frac{g'_{\alpha 4}g'_{\beta 4}}{g'_{44}}\right)\mathrm{d}x'^\alpha\mathrm{d}x'^\beta\,,\quad \mbox{where} \quad \alpha, \beta=1,2,3\,.
\end{equation}
Thus the line element $\mathrm{d}\sigma^2$ corresponds to the result of the ``radar method'' of Landau and Lifschitz \cite[§~84]{LL}. Since the projection tensor $h'_{\alpha\beta}$ does not depend on $x'^4$ in our case, we can also use it to calculate finite proper spatial distances via integration. This is due to the fact that the spacetime is stationary and the four-velocity field shares this symmetry. %(although it is not hypersurface-orthogonal).
However, the four-velocity field is not hypersurface-orthogonal and the 3-space characterized globally by the line element (\ref{eq:LL}) is not a hypersurface of spacetime.

From now on we are only interested in the proper spatial geometry of the disc  itself, as a 2-dimensional object - described by the line element  \mbox{$\mathrm{d}\sigma^2\Big\vert_{\text{disc}}=\left\{h'_{\alpha\beta}\,\mathrm{d}x'^\alpha\mathrm{d}x'^\beta\right\}\!\Big\vert_{\text{disc}}$}. This is the geometry observed by a family of ``residents'' of  the disc who apply the radar method.

In elliptic coordinates the 2-dimensional proper spatial line element of the metric (\ref{eq:bvp.4}) takes the form
%On the disc itself, the 2-dimensional proper spatial line element of the metric (\ref{eq:bvp.4}) in elliptic coordinates takes the form
%By using the radar method of Landau and Lifschitz, the 2-dimensional proper spatial line element of the metric (\ref{eq:bvp.4}) in elliptic coordinates takes the form
\begin{equation} \label{eq:rep.1}
	\mathrm{d}\sigma^2 \Big\vert_{\nu=0} = f'^{-1}h'\rho_0^2\frac{\eta^2}{1-\eta^2}\mathrm{d}\eta^2 + f'^{-1}\rho_0^2(1-\eta^2)\mathrm{d}\varphi'^2 \,.
\end{equation}
Note that, since $\nu=0$ on the disc, the elliptic coordinates decouple and $\mathrm{d}\rho^2=\rho_0^2\frac{\eta^2}{1-\eta^2}\mathrm{d}\eta^2$ and $\rho^2=\rho_0^2(1-\eta^2)$.

Integration of the line element (\ref{eq:rep.1}) then gives the  %\newline
proper circumference
\begin{equation}\label{eq:rep.2}
	C'(\eta) = 2\pi \rho_0\sqrt{1-\eta^2}f'^{-1/2} \,,
\end{equation}
the proper area
\begin{equation}\label{eq:rep.2.5}
	A'(\eta) = 2\pi \rho_0^2 \int_\eta^1\!\mathrm{d}\tilde{\eta} \,\tilde{\eta}\, f'^{-1}h'^{1/2} 
\end{equation} 
and the proper radius
\begin{equation}\label{eq:rep.3}
	R(\eta) = R'(\eta) = \rho_0 \int_\eta^1\!\mathrm{d}\tilde{\eta} \,\frac{\tilde{\eta}}{\sqrt{1-\tilde{\eta}^2}}\left(f'^{-1}h'\right)^{1/2} 
\end{equation}
of the rotating disc of charged dust. %\footnote{The condition of positive definiteness to the proper spatial line element is trivially fulfilled here, since (proper) radius and (proper) circumference are positive definite.}
%\footnote{The proper spatial line element has to be positive definite, which is fulfilled trivially here, since radius and circumference are positive definite. Furthermore, integration of the line element is only possible due to the time-independent metric.}
By inserting the post-Newtonian expansions of the metric functions in the co-rotating frame, using (\ref{eq:pne.expansions}), (\ref{eq:pne.3}) and (\ref{eq:bvp.5}), also (\ref{eq:rep.2}), (\ref{eq:rep.2.5}) and (\ref{eq:rep.3}) can be written as series expansions in $g$.\footnote{When we state that the geometry is ``seen'' or ``measured'' by co-rotating observers, we actually mean that the co-rotating observers apply the radar method to measure distances. Since they are at rest relative to the rotating disc (and their four-velocity field, in accordance with the four-velocity field of fixed particles in the disc, has only a fourth component in the co-rotating frame), they measure the proper spatial geometry of the disc. In the non-rotating frame, on the other hand, the non-rotating observers also use the radar method. However, they are at rest relative to the non-rotating frame (with a four-velocity field that has vanishing spatial components in the non-rotating frame) and consequently do not measure the proper spatial geometry of the disc, but the proper spatial geometry of their own ``rest space''. \\
Accordingly, in the non-rotating frame of reference circumference and radius are defined by \mbox{$C(\eta) = 2\pi \rho_0\sqrt{1-\eta^2}f^{-1/2}$} and $R(\eta) = \rho_0 \int_\eta^1\!\mathrm{d}\tilde{\eta} \,\frac{\tilde{\eta}}{\sqrt{1-\tilde{\eta}^2}}\left(f^{-1}h\right)^{1/2}$, respectively.}

%\footnote{When we state that the geometry is ``seen" or ``measured" by co-rotating observers, we actually mean that the co-rotating observers apply the radar method using ``light rulers" to measure distances. Then (but generally not otherwise) the co-rotating observers measure the proper spatial geometry of the disc. Non-rotating observers, on the other hand, also use ``light rulers" to measure distances. However, since the measurements are performed in the non-rotating frame, they do not measure the proper spatial geometry of the disc.}\footnote{In general, $u^i$ is the four-velocity field of particles or observers at rest relative to the appropriate reference frame.}\footnote{Accordingly, in the non-rotating frame of reference circumference and radius are defined by \mbox{$C(\eta) = 2\pi \rho_0\sqrt{1-\eta^2}f^{-1/2}$} and $R(\eta) = \rho_0 \int_\eta^1\!\mathrm{d}\tilde{\eta} \,\frac{\tilde{\eta}}{\sqrt{1-\tilde{\eta}^2}}\left(f^{-1}h\right)^{1/2}$, respectively.}

In order to compare the results for $C'$, $A'$ and $R$ of the charged disc of dust with those of a standard disc within the framework of special relativity, described by Gr\o{}n, we have to take the Newtonian limit. 
Thus, we require that both the metric deviates only slightly from the Minkowski metric and the rotational velocities are small compared to the speed of light. 
%Therefore, we demand that both the metric to only deviate slightly from the Minkowski metric and rotational velocities that are small compared to the speed of light. 
%This way we can compare the findings for $C'$, $A'$ and $R$ with the ones of a standard disc within the framework of special relativity described by Gr\o{}n. 
\\
%\begin{minipage}[t]{0.55\textwidth}
%	Charged disc of dust (Newtonian limit):
%	\begin{align*}
%		&C' = 2\pi\rho_0\left[{\rho^{\star}} + \frac{1}{2}\left({\rho^{\star}}^{3} {\Omega_{1}^{\star}}^{2}-{\rho^{\star}} f_{2}\right) g^{2}\right] \,,\\
%		&A' = \pi\rho_0^2\left[{\rho^{\star}}^2 + \left(\int_{\eta}^{1} \tilde{\eta}  \left({\Omega_{1}^{\star}}^{2} {\rho^{\star}}^{2}-2 f_{2}\right)\!\mathrm{d} \tilde{\eta}\! \right) \!g^{2}\right] \,,\\
%		&R = \rho_0\left[{\rho^{\star}} - \frac{1}{2}\left(\int_{\eta}^{1}\frac{\tilde{\eta}  f_{2}}{\sqrt{1-\tilde{\eta}^{2}}}\mathrm{d}\tilde{\eta} \right) g^{2}\right] \,. \\
%		%&w_{\varphi'} = 1-{\Omega_{1}^{\star}} {\rho^{\star}} g +\frac{1}{2} \left({\rho^{\star}}^{2} {\Omega_{1}^{\star}}^{2} + f_{2} \right) g^{2} 
%	\end{align*}
%%	where 
%%	\begin{equation*}
%%		\rho^{\star} = \frac{\rho}{\rho_{0}} = \sqrt{1-\eta^{2}} \,, \quad f_{2}\! \left(\eta,0\right) = -1-\eta^{2}
%%	\end{equation*}
%\end{minipage} 
%\hfill
%\begin{minipage}[t]{0.33\textwidth}
%		Disc within SR (Newtonian limit):
%		\begin{align*}
%			&C' = 2\pi r\left[1 +  
%			\frac{1}{2}\left(\omega r\right)^2\right] \,, \\
%			&A' = \pi r^2\left[1+ \frac{1}{4}\left(\omega r\right)^2 \right] \,, \\
%			&r = r \,. \\
%			%&w_{\varphi'} = 1 - \,{\omega r}\, + 
%			%\frac{1}{2}\left(\omega r\right)^2
%	\end{align*} 
%\end{minipage}\\
\begin{minipage}[t]{0.55\textwidth}
Charged disc of dust (Newtonian limit):
\begin{align*}
	&C' = 2\pi\rho_0\left[{\rho^{\star}} + \left( \frac{1}{2}{\Omega_{1}^{\star}}^{2}{\rho^{\star}}^{2}- \frac{1}{2}f_{2}\right) {\rho^{\star}}g^{2}\right] \,,\\
	&A' = \pi\rho_0^2\left[{\rho^{\star}}^2 + \left( \frac{1}{4}{\Omega_{1}^{\star}}^{2}{\rho^{\star}}^{2} + \frac{1}{2}\left(2-f_{2}\right) \right) \!{\rho^{\star}}^{2}g^{2}\right] \,,\\
	&R = \rho_0\left[{\rho^{\star}} + \frac{1}{6}\left( 4-f_{2} \right) \rho^{\star}g^{2}\right] \,.
	%&w_{\varphi'} = 1-{\Omega_{1}^{\star}} {\rho^{\star}} g +\frac{1}{2} \left({\rho^{\star}}^{2} {\Omega_{1}^{\star}}^{2} + f_{2} \right) g^{2} 
\end{align*}
%	where 
%	\begin{equation*}
%		\rho^{\star} = \frac{\rho}{\rho_{0}} = \sqrt{1-\eta^{2}} \,, \quad f_{2}\! \left(\eta,0\right) = -1-\eta^{2}
%	\end{equation*}
\end{minipage} 
\hfill
\begin{minipage}[t]{0.33\textwidth}
Disc within SR (Newtonian limit):
\begin{align*}
	&C' = 2\pi r\left[1 +  
	\frac{1}{2}\left(\omega r\right)^2\right] \,, \\
	&A' = \pi r^2\left[1+ \frac{1}{4}\left(\omega r\right)^2 \right] \,, \\
	&R = r \,.
	%&w_{\varphi'} = 1 - \,{\omega r}\, + 
	%\frac{1}{2}\left(\omega r\right)^2
\end{align*} 
\end{minipage}
\vspace{0.5cm}
\newline
Crucial to the discussion of Ehrenfest's paradox and the closely related geometric questions is the ratio of circumference to radius as seen from co-rotating and non-rotating observers. \\
%\begin{minipage}[t]{0.55\textwidth}
%	Charged disc of dust (Newtonian limit): 
%	\begin{align*}
%		&\frac{C'}{R'} = 2\pi\Bigg[1 + \frac{1}{2}\Bigg(\frac{1}{\rho^{\star}}\int_{\eta}^{1}\frac{\tilde{\eta}f_{2}}{\sqrt{1-\tilde{\eta}^{2}}}\mathrm{d}\tilde{\eta} \\
%		&\hspace{2.85cm}- f_{2} +{\Omega_{1}^{\star}}^{2} {\rho^{\star}}^2\Bigg)\,g^{2}\Bigg] \,, \\
%		&\frac{C}{R} = 2\pi\left[1+\frac{1}{2}\left(\frac{1}{{\rho^{\star}}}\int_{\eta}^{1}\frac{\tilde{\eta}f_{2}}{\sqrt{1-\tilde{\eta}^{2}}}\mathrm{d}\tilde{\eta} - f_{2}\right) \,g^{2}\right] \,. \\
%	\end{align*}
%\end{minipage} 
%\hfill
%\begin{minipage}[t]{0.33\textwidth}
%	Disc within SR (Newtonian limit):
%	\begin{align*}
%		&\frac{C'}{r} = 2\pi \left[1 + \frac{1}{2}\left(\omega r\right)^2\right] \\
%		&\hspace{0.5cm} > 2\pi \,, \\
%		&\frac{C}{r}= 2\pi \,. \\
%\end{align*} 
%\end{minipage}
\begin{minipage}[t]{0.55\textwidth}
Charged disc of dust (Newtonian limit): 
\begin{align*}
	&\frac{C'}{R'} = 2\pi\left[1 + \left( \frac{1}{2}{\Omega_{1}^{\star}}^{2}{\rho^{\star}}^{2} - \frac{1}{3}\left(2+f_{2}\right)\right)g^2 \right] \,, \\
	&\frac{C}{R} = 2\pi\left[1  - \frac{1}{3}\left(2+f_{2}\right)g^2 \right] \,.
\end{align*}
\end{minipage} 
\hfill
\begin{minipage}[t]{0.33\textwidth}
Disc within SR (Newtonian limit):
\begin{align*}
	&\frac{C'}{r} = 2\pi \left[1 + \frac{1}{2}\left(\omega r\right)^2\right] \,, \\
	%&\hspace{0.5cm} > 2\pi \,, \\
	&\frac{C}{r}= 2\pi \,.
\end{align*} 
\end{minipage}
\vspace{0.5cm}
\newline
The first-order ($n=1$) coefficient function $f_2$ is given by $f_{2}\! \left(\eta,0\right) = -1-\eta^{2}$ and $\rho^{\star} = \frac{\rho}{\rho_{0}} = \sqrt{1-\eta^{2}}$.

It should be noted that the above expansions utilize different expansion parameters and therefore only in the Newtonian limit a comparison is meaningful. For the disc of dust we demand the gravitational potential to be small compared to the speed of light $c$ and the expansion parameter is $g=\frac{\sqrt{-U_c}}{c} \ll 1$. On the other hand, for the SR-disc we have expansions around small rotational velocities, $\frac{\omega r}{c} \ll 1$.

Gr\o{}n discusses only kinematic aspects of the disc and material properties are absent in \cite{Gron1975, Gron1977}. However, a real disc is hold together by attractive forces originating from the rigid material itself. In case of the disc of dust attractive gravitational forces, mediated through the terms involving $f_{2}$, play the role of those material forces.

Taking the different expansions and the appearance of the terms with $f_{2}$ into account, the disc of dust agrees with Gr\o{}n's investigated disc in the Newtonian limit.

Furthermore, an interesting observation can be made. The quantities $C', A'$ and $R$ of the disc of dust are all larger than those of Gr\o{}n's analysed disc (for $\eta^2<1$ and $r>0$). Although dust is by definition a pressure-less fluid without any elastic properties, the rotating disc of dust nevertheless behaves to some extend as if it had material properties, resulting in a kind of ``elastic'' expansion.

\begin{figure}
	\centering
	\begin{minipage}{0.45\textwidth}
		\centering
		\begin{tikzpicture}
			\node[anchor=south west, inner sep=0] (image) at (0,0) {\includegraphics[width=1\textwidth, trim={3.5cm 17.5cm 5.5cm 5.5cm}, clip]{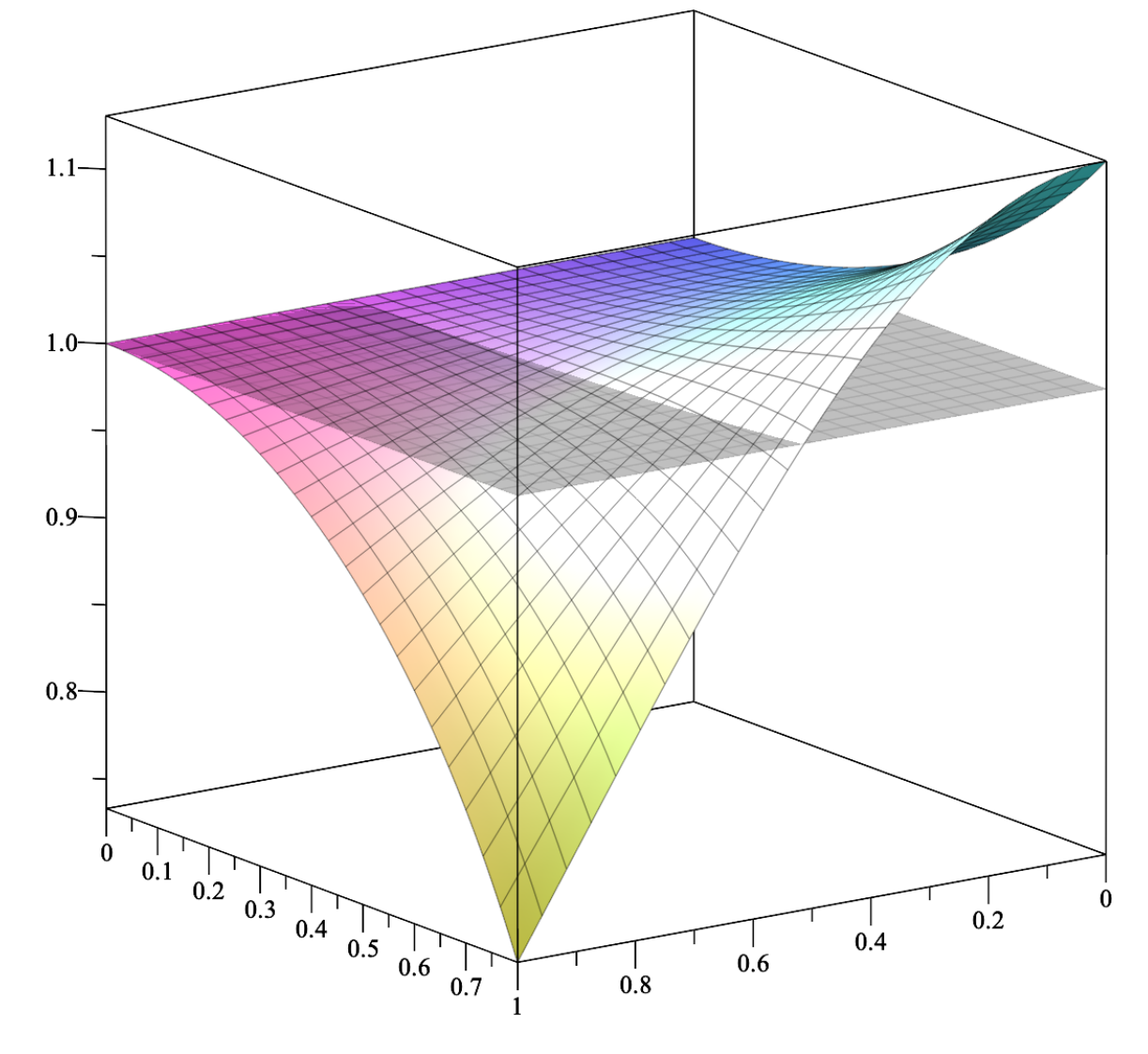}};
			\begin{scope}[x={(image.south east)}, y={(image.north west)}]
				\node [anchor=east] at (0.04, 0.567) {$\frac{1}{2\pi}\frac{C'}{R'}$};
				\node at (0.273, 0.05) {$g$};
				\node at (0.713 ,0.05) {$\epsilon$};
			\end{scope}
		\end{tikzpicture}
		\caption{Normalized ratio of circumference to radius in the rotating frame plotted for $\eta=0$.}
		\label{fig:1}
	\end{minipage}\hfill
	\begin{minipage}{0.45\textwidth}
		\centering
		\begin{tikzpicture}
			\node[anchor=south west, inner sep=0] (image) at (0,0) {\includegraphics[width=1\textwidth, trim={3.5cm 17.5cm 5.5cm 5.5cm}, clip]{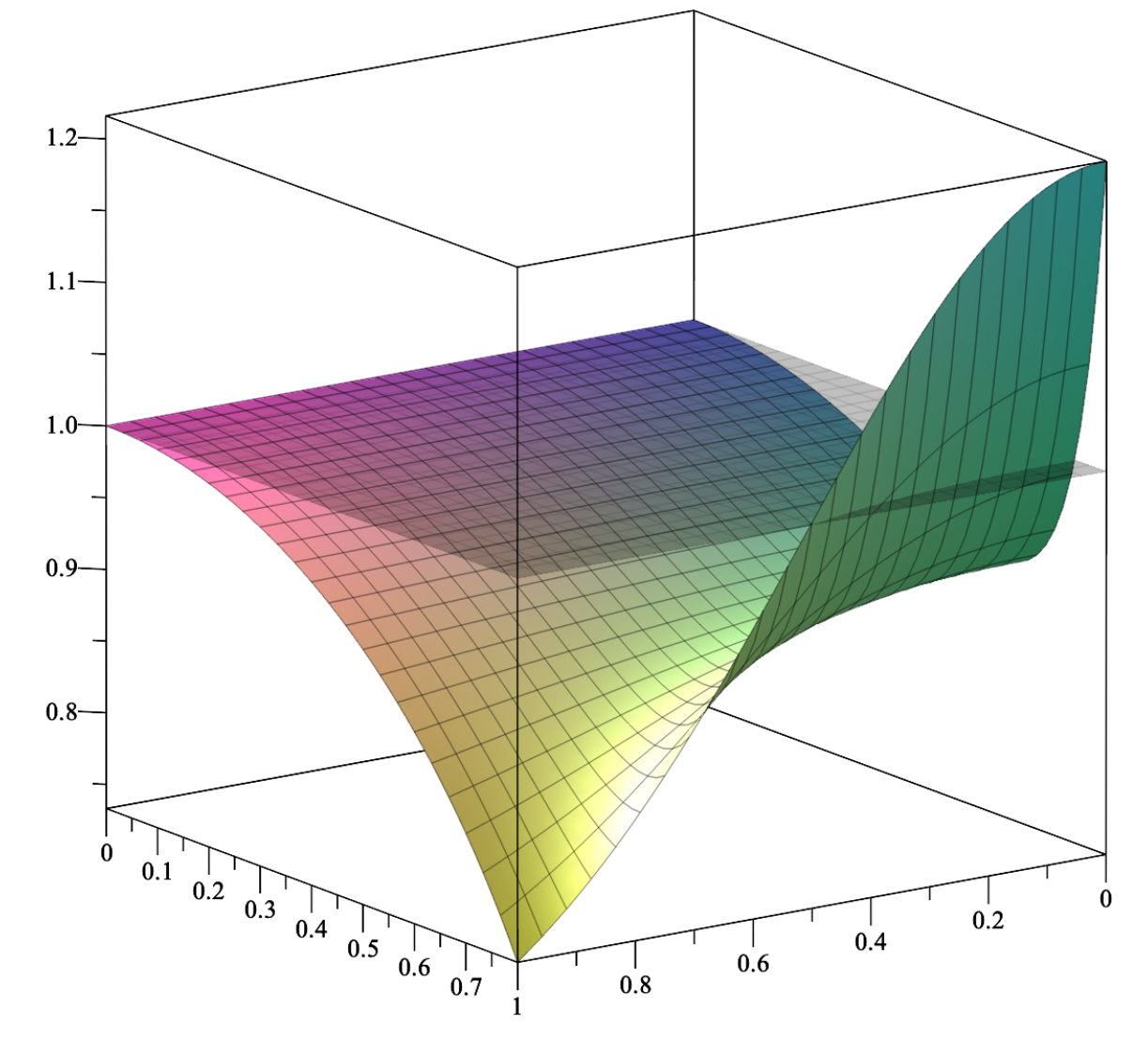}};
			\begin{scope}[x={(image.south east)}, y={(image.north west)}]
				\node[anchor=east] at (0.04, 0.567) {$\frac{1}{2\pi}\frac{C}{R}$};
				\node at (0.273, 0.05) {$g$};
				\node at (0.713 ,0.05) {$\epsilon$};
			\end{scope}
		\end{tikzpicture}
		\caption{Normalized ratio of circumference to radius in the non-rotating frame plotted for $\eta=0$.}
		\label{fig:2}
	\end{minipage}
\end{figure}

Explicitly writing out the ratios of circumference to radius leads to:
\vspace{-0.25cm}
\begin{align}
	%&C' = 2\pi\rho_0\left[\sqrt{1-\eta^{2}}+\frac{1}{2}\left(2-\left(1-\eta^{2}\right) \epsilon^{2}\right) \sqrt{1-\eta^{2}}\, g^{2}\right] \,, \\
	%&A' = \pi {\rho_0}^2\left[\left(1-\eta^2\right) + \frac{1}{4}\left(8- \left(1+\epsilon^2\right) \!\left(1-\eta^2\right) \right) \left(1-\eta^2\right) g^{2} \right] \,, \\
	%&R = \rho_0\left[\sqrt{1-\eta^{2}}+\frac{1}{6}\sqrt{1-\eta^{2}}\, \left(\eta^{2}+5\right) g^{2} \right] \,, \\
	&\frac{C'}{R'} = 2\pi\left[1+ \frac{1}{6}\left(1-\eta^2\right)\!\left(1-3\epsilon^2\right) g^{2}  \right] \,, \\
	&\frac{C}{R} = 2\pi\left[1-\frac{1}{3}\left(1-\eta^{2}\right) g^{2} \right] \,.
\end{align}
Remarkably, for $\eta^2<1$, there is a geometrical transition induced by an alteration of the specific charge: $\frac{C'}{R'} \gtrless 2\pi$ for $\epsilon^2 \lessgtr \frac{1}{3}$. This transition is also present beyond the Newtonian limit, as can be seen in figure \ref{fig:1}. All figures in this paper are created using all available orders, i.e.\ up to $n=10$. For vanishing charge the ratio of proper circumference to proper radius becomes larger than $2\pi$, just as in the case of Gr\o{}n's result for this ratio.

In the non-rotating frame there is no such transition, instead the ratio of circumference to radius is always smaller than $2\pi$, independent of the specific charge (excluding $\eta=\pm1$). This also stays true beyond the Newtonian limit (apart from high values of $g$), see figure \ref{fig:2}. Within special relativity the ratio is exactly equal to $2\pi$, as Gr\o{}n showed. Those results, however, do not contradict each other, since values smaller than $2\pi$ are caused by the gravitational correction term involving $f_{2}$. As discussed above, Gr\o{}n's analysis of the rotating disc does not involve corresponding terms.\footnote{In the framework of special relativity there is an enlightening explanation for Gr\o{}n's results: Observed in the non-rotating frame, not the circumference of the disc is being contracted due to rotation, but the infinitesimal measuring rods placed along the circumference. As a consequence observers on the rotating disc measure a ratio larger than $2\pi$ (they have to use more measuring rods), while an observer in the non-rotating frame still measures exactly $2\pi$.}

%\begin{figure}[htb]
%	\centering
%	\includegraphics[width=0.6\textwidth, trim={3.5cm 17.5cm 5.5cm 5.5cm}, clip]{pictures/test4}
%	\caption{test}\label{fig:1}
%\end{figure}

%\begin{figure}[htb]
%	\centering
%	\begin{tikzpicture}
%		\node[anchor=south west, inner sep=0] (image) at (0,0) {\includegraphics[width=0.5\textwidth, trim={3.5cm 17.5cm 5.5cm 5.5cm}, clip]{pictures/UR_rot}};
%		\begin{scope}[x={(image.south east)}, y={(image.north west)}]
%		\node at (-0.01, 0.567) {$\frac{C'}{R'}$};
%		\node at (0.273, 0.05) {$g$};
%		\node at (0.713 ,0.05) {$\epsilon$};
%		\end{scope}
%	\end{tikzpicture}
%	\caption{Ratio circumference to radius in rotating frame}
%	\label{fig:1}
%\end{figure}
%\begin{figure}[htb]
%	\centering
%	\begin{tikzpicture}
%		\node[anchor=south west, inner sep=0] (image) at (0,0) {\includegraphics[width=0.5\textwidth, trim={3.5cm 17.5cm 5.5cm 5.5cm}, clip]{pictures/UR}};
%		\begin{scope}[x={(image.south east)}, y={(image.north west)}]
%			\node at (-0.01, 0.567) {$\frac{C}{R}$};
%			\node at (0.273, 0.05) {$g$};
%			\node at (0.713 ,0.05) {$\epsilon$};
%		\end{scope}
%	\end{tikzpicture}
%	\caption{Ratio circumference to radius in non-rotating frame}
%	\label{fig:2}
%\end{figure}

Now, let us address the questions raised in the beginning of this section, starting with the first one. 
Formulated in the Newtonian language, a dust particle in the rotating disc is in an equilibrium state of gravitational, electric and centrifugal force. It is therefore evident that changing the parameter $\epsilon$ directly affects the rotational velocity $\Omega$. In particular, by decreasing $\epsilon$ we can transition to disc configurations with higher rotational velocities $\Omega$. This, however, is achieved in a quasi-stationary way and cannot be seen as setting the disc into rotation as we alter the specific charge of the disc and thus the disc itself. Nevertheless, we can compare disc configurations with increasing rotational velocities with each other and examine the effect on the spatial geometry. As can be seen in \mbox{figure \ref{fig:1}}, for co-rotating observers the geometric ratio $\frac{C'}{R'}$ monotonically increases from values smaller to greater than $2\pi$. 
%In fact, as will be shown in section \ref{sec6}, the curvature of the disc simultaneously changes from positive to negative.
In fact, as will be shown in section \ref{sec6}, a simultaneous transition from positive to negative curvature also occurs.
Non-rotating observers also measure increasing values of $\frac{C}{R}$. They, however, remain below $2\pi$, apart from high values of $g$ due to the emergence of an ergosphere.\footnote{We note that inside an ergosphere non-rotating (with respect to infinity) observers no longer exist.}
%Apart from general relativistic effects at high values  of $g$ due to the emergence of an ergosphere, the non-rotating observer always measures $\frac{C}{R} < 2\pi$, though, with increasing numbers for growing $\Omega$, see figure \ref{fig:2}.

Concerning the second question, 
we have seen that the ratio $\frac{C'}{R'}$, that characterizes the proper spatial geometry and is measured by co-rotating observers, can be less than, equal to or greater than $2\pi$ depending on the specific charge $\epsilon$. For a ratio equal to $2\pi$ the critical value of the specific charge is $\epsilon_{\text{crit}}=\frac{1}{\sqrt{3}}$ in the Newtonian limit. With growing $g$, $\epsilon_{\text{crit}}$ decreases slightly and becomes dependent on the radial coordinate $\eta$ (see figures \ref{fig:10.1} and \ref{fig:10.2}).
Non-rotating observers, on the other hand, always perceive smaller values than $2\pi$ for $\frac{C}{R}$ (again, apart from high values of $g$).

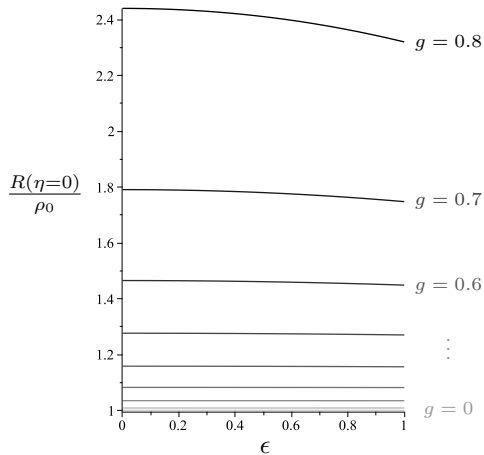
\begin{figure}[htb]
	\centering
	\begin{tikzpicture}
		\node[anchor=south west, inner sep=0] (image) at (0,0) {\includegraphics[width=0.5\textwidth]{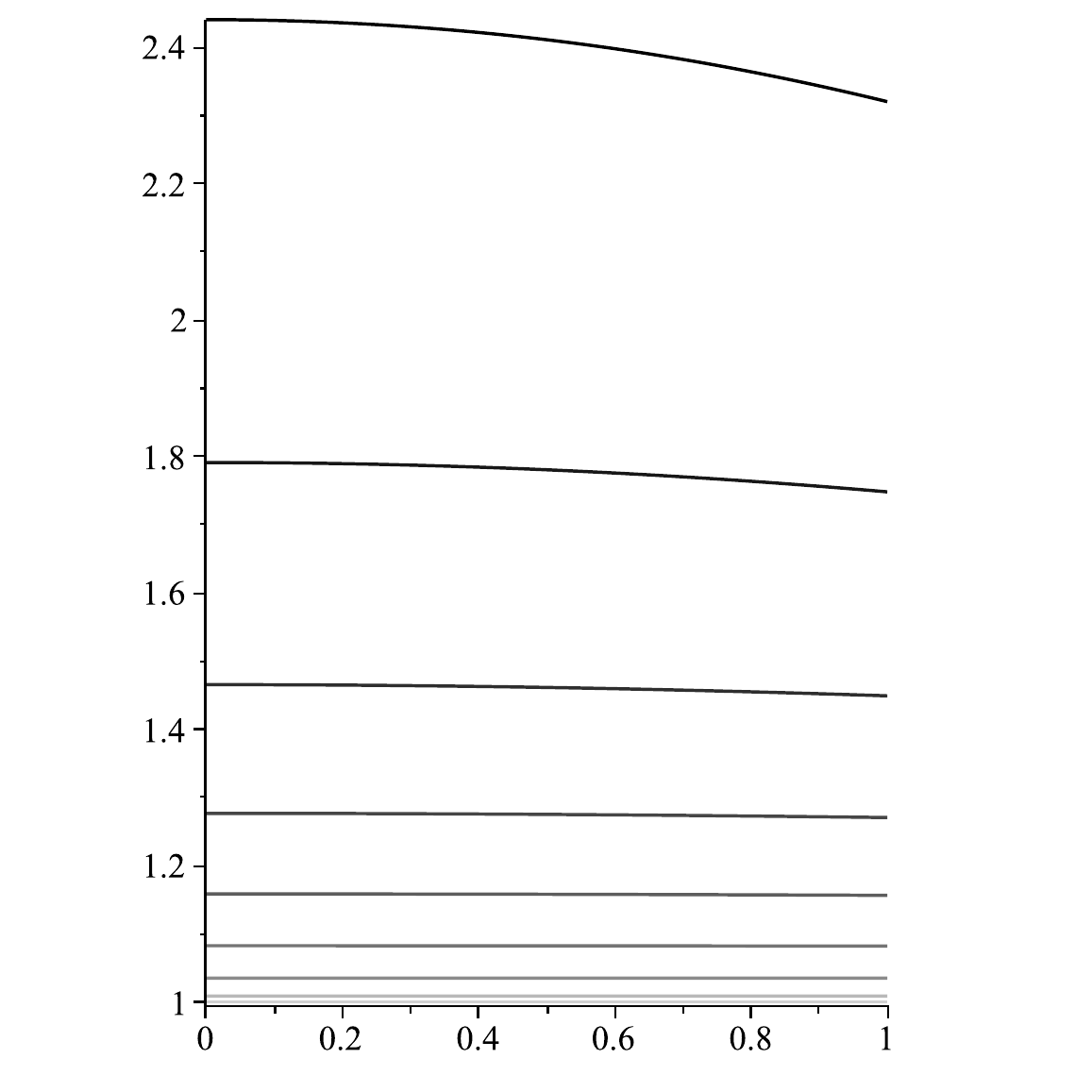}};
		\begin{scope}[x={(image.south east)}, y={(image.north west)}]
			\node[anchor=east] at (0.125, 0.567) {$\frac{R(\eta=0)}{\rho_{0}}$};
			\node at (0.505 ,0.0) {$\epsilon$};
			\node[color=black] at (0.91, 0.903) {\footnotesize $g=0.8$};
			\node[color=Black!90] at (0.91, 0.55) {\footnotesize $g=0.7$};
			\node[color=Black!80] at (0.91, 0.36) {\footnotesize $g=0.6$};
			\node[color=Black!60] at (0.91, 0.24) {\footnotesize $.$};
			\node[color=Black!60] at (0.91, 0.22) {\footnotesize $.$};
			\node[color=Black!60] at (0.91, 0.20) {\footnotesize $.$};
			\node[color=Black!50] at (0.91, 0.085) {\footnotesize $g=0$};
		\end{scope}
	\end{tikzpicture}
	\caption{Normalized proper radius, evaluated at the rim of the disc, as a function of $\epsilon$.}
	\label{fig:3}
\end{figure}

Finally, the third question resolves around the influence of material properties present in a real rotating disc. Indeed, as we have already seen in the Newtonian limit, the disc of dust mimics ``elastic'' material properties and shows an ``elastic'' expansion as response to the rotational motion. But also beyond Newtonian physics, according to figure \ref{fig:3}, the disc's total proper radius $R_{0} \coloneqq R(\eta=0)$ grows with decreasing specific charge $\epsilon$ and in turn increasing rotational velocities $\Omega$. The more relativistic the disc, the more pronounced is the effect.

\section{Curvature}\label{sec5}

Motivated by the found geometric transition induced by a change of the specific charge $\epsilon$, we now examine the intrinsic curvature of rotating discs more closely.

As the domain of a rotating disc is simply a two-dimensional spatial surface, denoted by $\Sigma_2$, the geometric quantity that describes its intrinsic curvature is the Gaussian curvature $K$.

In the following subsections we will calculate the Gaussian curvature of the charged rotating disc of dust (using the post-Newtonian expansion up to \mbox{$n=10$}) and various analytic limiting cases. The findings will be compared and reviewed.

\subsection{Gaussian curvature of the charged disc of dust}\label{subsec5.1}

By using the derived expressions for the proper radius $R$ and the proper circumference $C'$, (\ref{eq:rep.3}) and (\ref{eq:rep.2}), the proper spatial line element of the charged rotating disc of dust can be written in the following instructive and simple way:
\begin{equation}\label{eq:gc.11}
	\mathrm{d}\sigma^2\Big\vert_{\Sigma_2} = \mathrm{d}R^2 + \left(\frac{C'(R)}{2\pi}\right)^{\!2}\mathrm{d}\varphi'^2 \,,
\end{equation} 
where at least formally $C'(R)=C'(\eta(R))$.

With this condensed version of the proper spatial line element also the Gaussian curvature takes a very simply form, i.e.
\begin{equation}\label{eq:gc.1}
	K = -\frac{C'(R)_{,RR}}{C'(R)} \,.
\end{equation}
It thus follows immediately that the Gaussian curvature of the charged rotating disc of dust is determined primarily by the second derivative of the proper circumference $C'$ with respect to the proper radius $R$.

\begin{figure}[htb]
	\centering
	\begin{minipage}{0.45\textwidth}
		\centering
		\begin{tikzpicture}
			\node[anchor=south west, inner sep=0] (image) at (0,0) {\includegraphics[width=0.9\textwidth, trim={3.5cm 17.5cm 5.5cm 5.5cm}, clip]{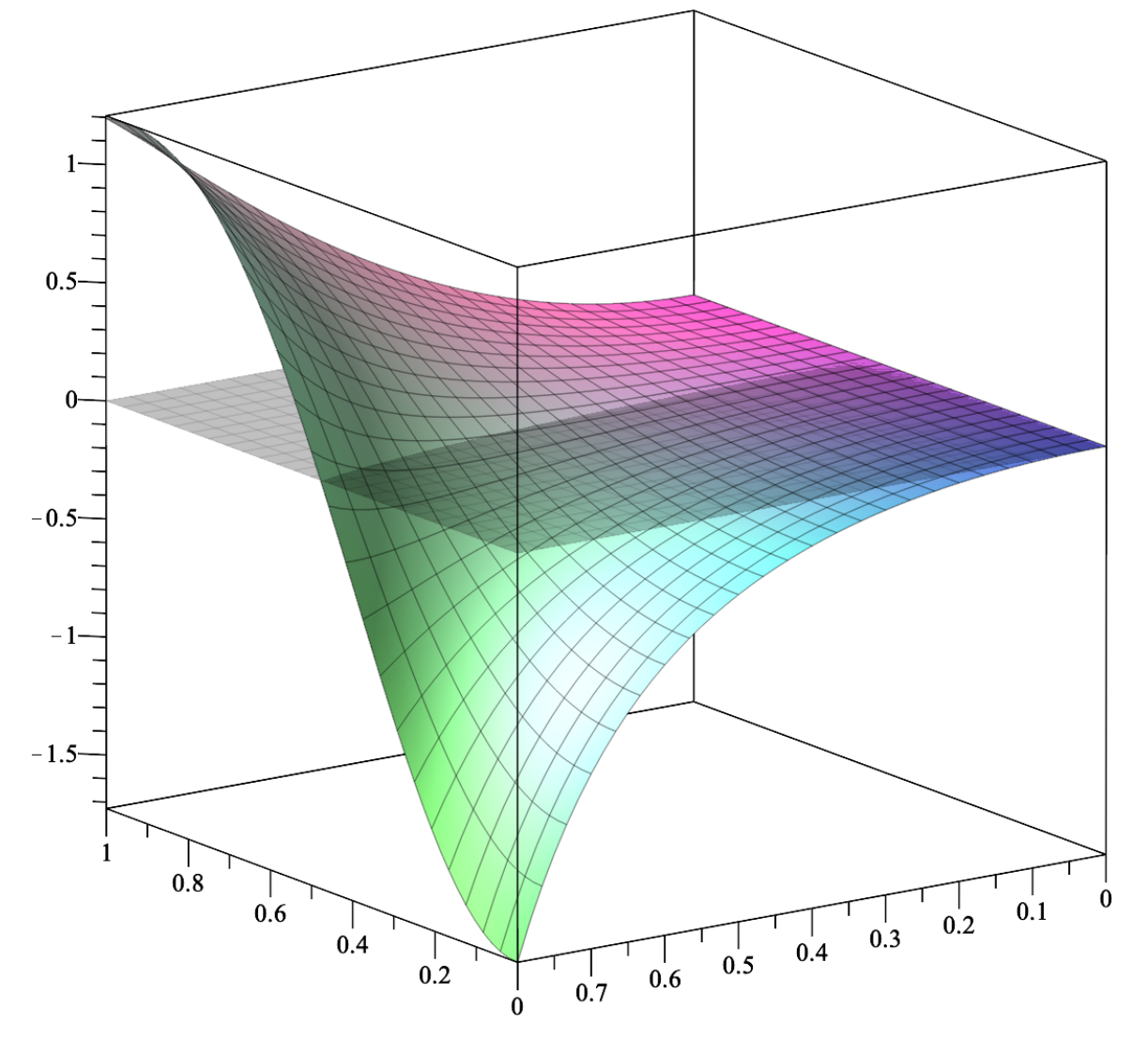}};
			\begin{scope}[x={(image.south east)}, y={(image.north west)}]
				\node[anchor=east] at (0.04, 0.567) {$R_{0}^2K$};
				\node at (0.273, 0.05) {$\epsilon$};
				\node at (0.713 ,0.036) {$g$};
			\end{scope}
		\end{tikzpicture}
		\caption{Gaussian Curvature $K$ normalized by the total proper radius, $R_{0}$, squared and evaluated at $\eta=0$.}
		\label{fig:4}
	\end{minipage}\hfill
	\begin{minipage}{0.45\textwidth}
		\centering
		\begin{tikzpicture}
			\node[anchor=south west, inner sep=0] (image) at (0,0) {\includegraphics[width=0.9\textwidth]{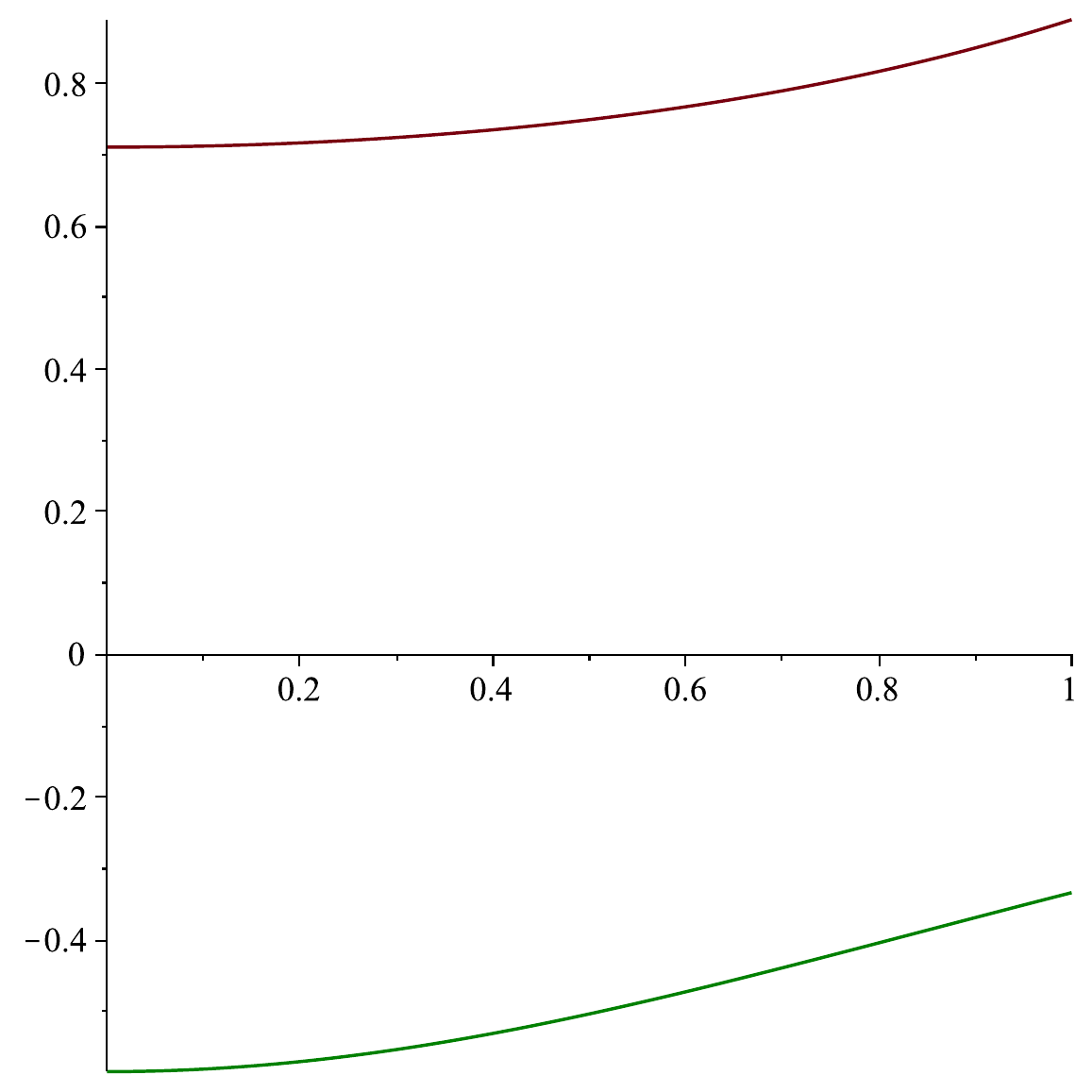}};
			\begin{scope}[x={(image.south east)}, y={(image.north west)}]
				\node[anchor=east] at (0.04, 0.567) {$R_{0}^2K$};
				\node at (0.543 ,0.32) {$\eta$};
				\node[color=BrickRed] at (0.8,0.875) {\footnotesize $\epsilon=1$};
				\node[color=OliveGreen] at (0.8,0.075) {\footnotesize $\epsilon=0$};
			\end{scope}
		\end{tikzpicture}
		\caption{Radial dependence of the normalized Gaussian curvature for the special cases $\epsilon=0$ and $\epsilon=1$, where $g=0.6$.}
		\label{fig:4.2}
	\end{minipage}
\end{figure}

Even though the above formula for $K$, (\ref{eq:gc.1}), is quite insightful, it is not that practical, since we do not know the functional dependence of $C'$ on $R$. By using the chain rule we can, however, transform formula (\ref{eq:gc.1}) into the slightly less appealing but all the more useful form
\begin{equation}
	K = \frac{1}{C'(\eta)}\left(\frac{\mathrm{d}R}{\mathrm{d}\eta}\right)^{\!-3}\left[ \frac{\mathrm{d}^2R}{\mathrm{d}\eta^2}C'(\eta)_{,\eta} - \frac{\mathrm{d}R}{\mathrm{d}\eta}C'(\eta)_{,\eta\eta} \right] \,.
\end{equation}
Inserting the solutions for the post-Newtonian expansion leads to
\begin{align}
	K = \frac{1}{\rho_0^2}\biggl[&\left(3 \epsilon^{2}-1\right) g^{2} \notag \\ 
%	&+ 6 \left(\left(\eta^{2}-\frac{5}{6}\right) \epsilon^{4}+\left(\frac{7}{36}-\frac{4 \eta^{2}}{3}\right) \epsilon^{2}+\frac{\eta^{2}}{2}+\frac{1}{12}+\frac{8 \eta^{2}}{3\pi^{2}}\left(\epsilon^2-1\right)\right) g^{4} \notag \\
	&+ \left(\left(6\eta^{2}-5\right) \epsilon^{4}+\left(\frac{7}{6}-8\eta^{2}\right) \epsilon^{2}+3\eta^{2}+\frac{1}{2}+\frac{16 \eta^{2}}{\pi^{2}}\left(\epsilon^2-1\right)\right) g^{4} \notag \\
	&+ \mathcal{O}\left(g^6\right)\biggr] \,.
\end{align}
As can be seen in figure \ref{fig:4}, there is a characteristic transition curve in parameter space where the curvature $K$ changes its sign. The same transition curve appears in figure \ref{fig:1} depicting $\frac{C'}{R'}$.
Explicitly written out expansions up to fifth order of $\frac{C'}{R'}$ and $R_{0}^2K$ are listed in appendix \ref{secA}.

Figure \ref{fig:4.2} shows the radial dependence of $R_{0}^2K$ for the special cases $\epsilon=1$ and $\epsilon=0$. In case of maximal charge, i.e.\ $\epsilon=1$, the disc has no rotation at all and general relativistic effects cause positive curvature. For $\epsilon=0$ the rotation is maximal and special relativistic effects are dominant and lead, in accordance with the disc described by Gr\o{}n, to negative curvature. In both cases the curvature $K$ gets more positive towards the centre of the disc (where $\eta=1$). This is expected, since in the non-rotating case general relativistic effects increase towards the centre of a massive object, where the gravitational potential is the deepest. At maximal rotation, on the other hand, the rotational speed decreases towards the centre while general relativistic effects increase.

\begin{figure}[htb]
	\centering
	\begin{minipage}{0.3\textwidth}
		\centering
		\begin{tikzpicture}
			\node[anchor=south west, inner sep=0] (image) at (0,0) {\includegraphics[width=1\textwidth]{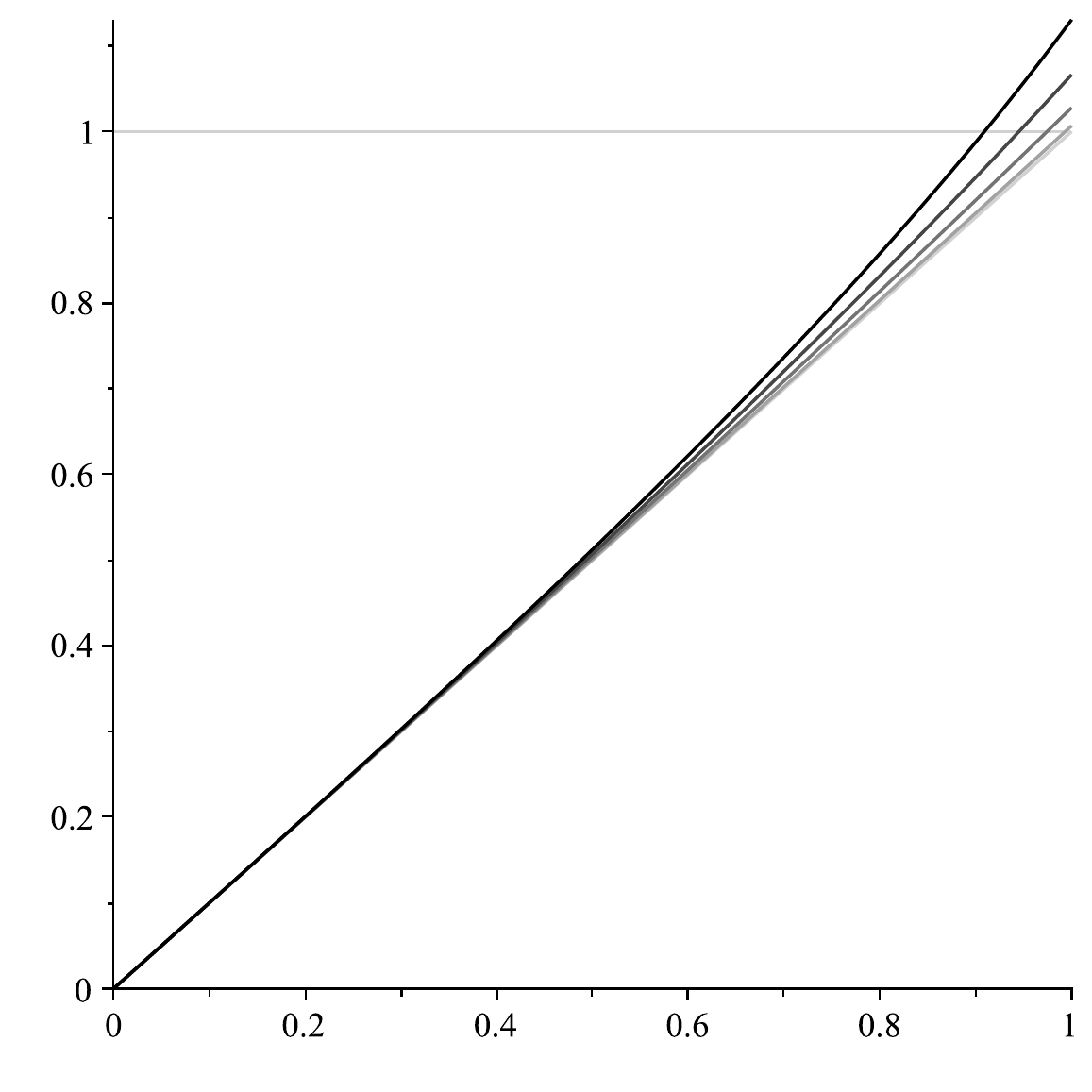}};
			\begin{scope}[x={(image.south east)}, y={(image.north west)}]
				\node[anchor=west] at (0.084, 0.567) {$\frac{C'}{2\pi R_{0}}$};
				\node at (0.543 ,-0.02) {$\frac{R}{R_{0}}$};
				\node[color=Black]  at (0.643 ,0.77) {\footnotesize $g=0.8$};
				\node[color=Black!50]  at (0.86 ,0.64) {\footnotesize $g=0$};
			\end{scope}
		\end{tikzpicture}
		\caption{Functional dependence of $C'$ (normalized by $2\pi R_{0}$) on $R$ (normalized by $R_{0}$) plotted for \mbox{$\epsilon=0$} and \mbox{$g=\{0, 0.2, 0.4, 0.6, 0.8\}$}.}
		\label{fig:5.1}
	\end{minipage}\hfill
	\begin{minipage}{0.3\textwidth}
		\centering
		\begin{tikzpicture}
			\node[anchor=south west, inner sep=0] (image) at (0,0) {\includegraphics[width=1\textwidth]{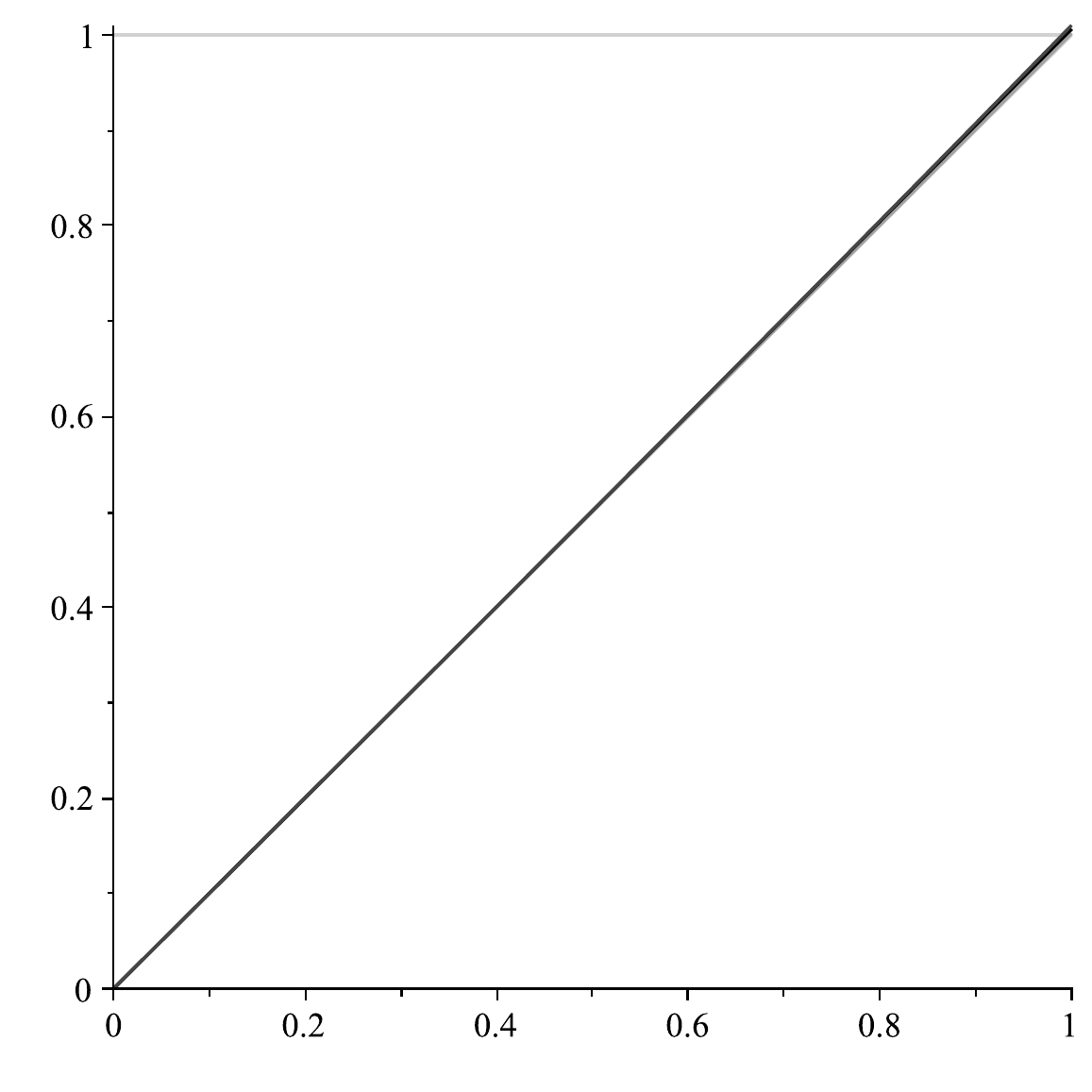}};
			\begin{scope}[x={(image.south east)}, y={(image.north west)}]
				\node[anchor=west] at (0.084, 0.567) {$\frac{C'}{2\pi R_{0}}$};
				\node at (0.543 ,-0.02) {$\frac{R}{R_{0}}$};
				\node[color=Black]  at (0.618 ,0.77) {\footnotesize $g=0.8$};
				\node[color=Black!50]  at (0.853 ,0.69) {\footnotesize $g=0$};
			\end{scope}
		\end{tikzpicture}
		\caption{Functional dependence of $C'$ (normalized by $2\pi R_{0}$) on $R$ (normalized by $R_{0}$) plotted for \mbox{$\epsilon=0.5$} and \mbox{$g=\{0, 0.2, 0.4, 0.6, 0.8\}$}.}
		\label{fig:5.2}
	\end{minipage}\hfill
	\begin{minipage}{0.3\textwidth}
		\centering
		\begin{tikzpicture}
			\node[anchor=south west, inner sep=0] (image) at (0,0) {\includegraphics[width=1\textwidth]{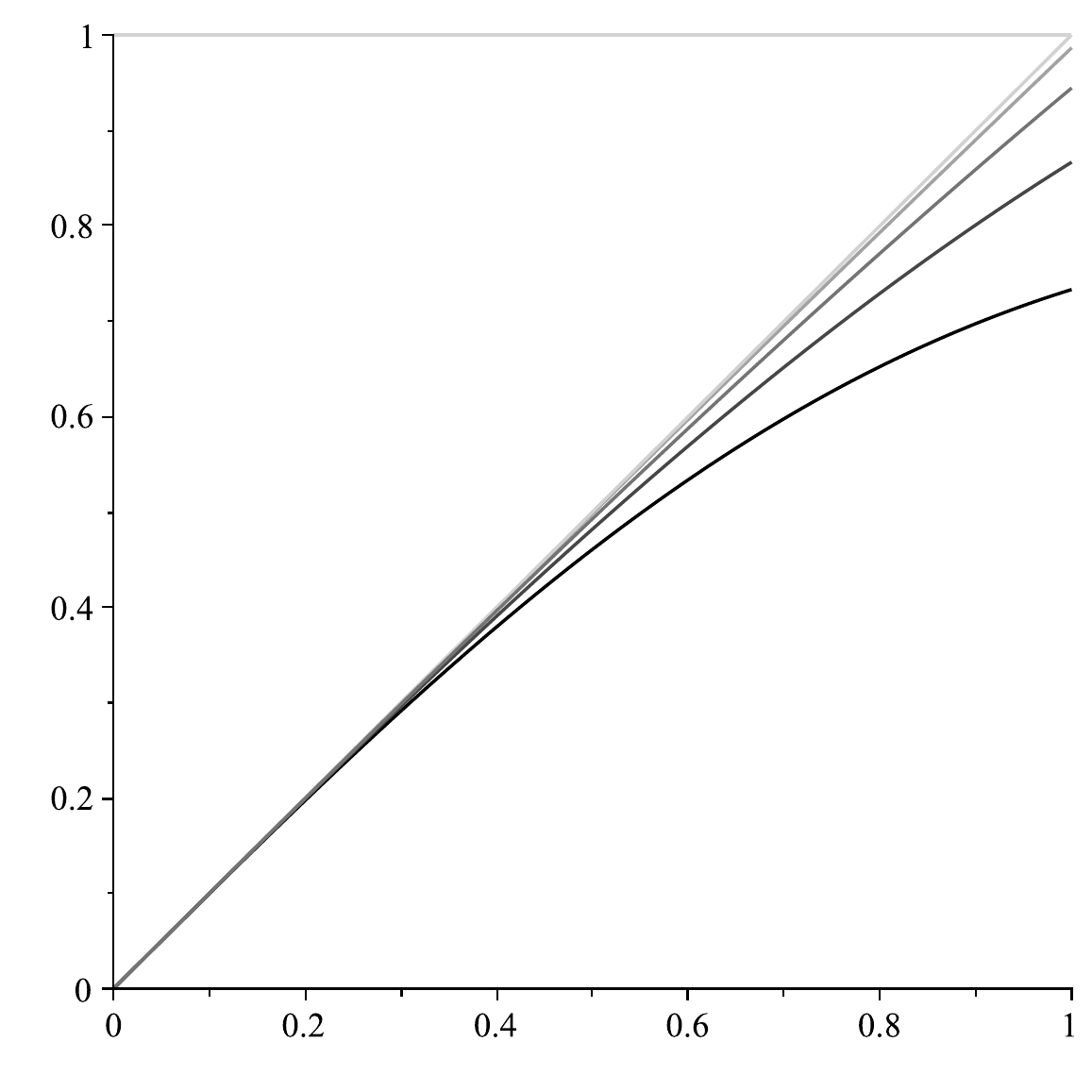}};
			\begin{scope}[x={(image.south east)}, y={(image.north west)}]
				\node[anchor=west] at (0.084, 0.567) {$\frac{C'}{2\pi R_{0}}$};
				\node at (0.543 ,-0.02) {$\frac{R}{R_{0}}$};
				\node[color=Black!50]  at (0.645 ,0.77) {\footnotesize $g=0$};
				\node[color=Black]  at (0.855 ,0.577) {\footnotesize $g=0.8$};
			\end{scope}
		\end{tikzpicture}
		\caption{Functional dependence of $C'$ (normalized by $2\pi R_{0}$) on $R$ (normalized by $R_{0}$) plotted for \mbox{$\epsilon=1$} and \mbox{$g=\{0, 0.2, 0.4, 0.6, 0.8\}$}.}
		\label{fig:5.3}
	\end{minipage}
\end{figure}

Interesting to the current discussion is also the functional dependence of $C'$ on $R$ that is visualized in terms of parametric plots (including appropriate normalizations) in figures \ref{fig:5.1} - \ref{fig:5.3}.
As can be observed there, $C'_{,RR}>0$ for small $\epsilon$ and $C'_{,RR}<0$ for large $\epsilon$. Considering formula (\ref{eq:gc.1}), this perfectly coincides with the plot of the normalized curvature in figure \ref{fig:4}.
The plots in figures \mbox{\ref{fig:5.1} - \ref{fig:5.3}} are also in line with the $\frac{C'}{R'}$-plot in figure \ref{fig:1}.

\subsection{Newtonian theory: Maclaurin discs and their Gaussian curvature}\label{subsec5.2}

In the Newtonian limit Einstein and Maxwell equations decouple and reduce to the Poisson equations for the gravitational and the electric potential. 
Thus, in Newtonian theory the disc of dust is fully described by
\begin{align}
	&\Delta U = 4\pi \mu^{\text{Mld}} \,, \quad \text{with} \quad \mu^{\text{Mld}} = \begin{cases} \sigma^{\text{Mld}}(\rho)\delta(\zeta)  &\text{for} \,\,\, 0\leq\rho\leq\rho_0 \\ 0 &\text{otherwise} \end{cases} \,, \label{eq:gc.2} \\
	&\Delta U^{\text{E}} = -4\pi \epsilon\mu^{\text{Mld}} \,, \label{eq:gc.3} 
\end{align}
and
\begin{equation} \label{eq:gc.4} 
	\epsilon {U^{\text{E}}}_{,\rho} = - \left(U-\frac{1}{2}\Omega^2\rho^2\right)_{\!,\rho} \,,
\end{equation}
where $U$ is the gravitational and $U^{\text{E}}\coloneqq \alpha = -A_t$ the electric potential. Equation (\ref{eq:gc.4}) shows that the dust particles in the disc are in an equilibrium of electric, gravitational and centrifugal force.

Rotating discs in the framework of Newtonian theory characterized by \mbox{(\ref{eq:gc.2}) - (\ref{eq:gc.4})} are also known as charged Maclaurin discs.

The solution for the exterior Newtonian gravitational potential of Maclaurin spheroids, in terms of elliptic coordinates, can be found in \cite{RFE}. In the limit where the spheroid shrinks to a disc, this potential is given everywhere. As before, we are interested in the solution on the disc itself. On $\Sigma_2$, where $\nu=0$, the gravitational potential reduces to
\begin{equation}\label{eq:gc.4.5} 
	U\vert_{\nu=0} = \frac{1}{2}U_c\left(1+\eta^2\right) \,,
\end{equation}
with $U_c=-g^2$ in the Newtonian limit. According to equations (\ref{eq:gc.2}) and (\ref{eq:gc.3}), $U^{\text{E}}$ trivially follows from equation (\ref{eq:gc.4.5}): $U^\text{E}=-\epsilon U$.

Furthermore, using the solution for $U$, (\ref{eq:gc.4}) can be rewritten to the already known equation, 
%With the help of the solution for $U$, equation (\ref{eq:gc.4}) can be rewritten to the already known equation, 
\begin{equation}\label{eq:gc.4.6} 
	U_c=-\frac{\Omega^2\rho_0^2}{1-\epsilon^2} \,,
\end{equation}
that was used as a starting point for the post-Newtonian expansion. 

Integrating (\ref{eq:gc.2}) over an infinitesimal $\zeta$-interval and exploiting reflection symmetry reveals the surface mass density
\begin{equation}
	\sigma^{\text{Mld}} = \frac{1}{2\pi}U_{,\zeta}\Big\vert_{\zeta=0^+} = \frac{3M}{2\pi\rho_0^2}\eta \,,%\sqrt{1-\frac{\rho^2}{\rho_0^2}} \,,
\end{equation}
where $M$ is the mass of the disc.
It can be easily verified that 
\begin{equation}\label{eq:gc.4.7} 
	U' = U - \frac{1}{2}\Omega^2\rho^2
\end{equation} 
is the corotating potential in the Newtonian limit.

By understanding the Newtonian theory as a limit of the full general relativistic theory we can reuse the line element. The 4-dimensional line element in the Newtonian limit, evaluated on the disc, reads
\begin{align} \label{eq:gc.5}
	\mathrm{d}s^2\Big\vert_{\nu=0} = &\left(1-2U\right)\rho_0^2\frac{\eta^2}{1-\eta^2}\mathrm{d}\eta^2 + \left(1-2U\right)\rho_0^2\left(1-\eta^2\right)\mathrm{d}\varphi'^2 \notag \\ 
	&+ 2\Omega\rho_{0}^2\left(1-\eta^2\right)\mathrm{d}\varphi'\mathrm{d}t - (1+2U')\mathrm{d}t^2 \,.
\end{align}
Using equation (\ref{eq:LL}), the proper spatial line element of the Maclaurin disc follows immediately:
%Using equations (\ref{eq:LL}), (\ref{eq:gc.4.5}), (\ref{eq:gc.4.6}) and (\ref{eq:gc.4.7}), the proper spatial line element of the Maclaurin disc follows immediately:
%Due to the absence of an off-diagonal term, the proper spatial line element of the Maclaurin disc follows immediately:%\footnote{As can be easily verified, here the proper spatial line element is simply the spatial part of (\ref{eq:gc.5}).}:
\begin{align}\label{eq:gc.5.5}
	d\sigma^2\Big\vert_{\nu=0} = &\left[1+\left(1+\eta^2\right)g^2\right]\rho_0^2\frac{\eta^2}{1-\eta^2}\mathrm{d}\eta^2 \notag \\
	%&+ \left[1+\left(\left(1+\eta^2\right)+\left(1-\epsilon^2\right)\left(1-\eta^2\right)\right)g^2\right]\left(1-\eta^2\right)\mathrm{d}\varphi'^2 \\
	&+ \left[1+\left(2-\epsilon^2\left(1-\eta^2\right)\right)g^2\right]\rho_0^2\left(1-\eta^2\right)\mathrm{d}\varphi'^2 \,.
\end{align}
As expected, the Gaussian curvature of the Maclaurin disc resulting from (\ref{eq:gc.5.5}) is given by:
\begin{equation}
	K^{\text{Mld}} = \frac{1}{\rho_0^2}\left(3 \epsilon^{2}-1\right) g^{2} \,.
\end{equation}

In conclusion, the Gaussian curvature of the charged Maclaurin disc is in perfect agreement with the Gaussian curvature of the disc of dust in the Newtonian limit.

\subsection{Gaussian curvature of a specific ECD-disc configuration}\label{subsec5.3}

For general (not necessarily disc-) configurations with $\epsilon=\pm1$ one obtains electrically counterpoised dust (ECD), see, e.g., \cite{Meinel_2011}. Those solutions are static. 
The Papapetrou-Majumdar class \cite{Papapetrou, Majumdar} contains such static solutions to the Einstein-Maxwell equations and the corresponding line element is of the form
\begin{equation}\label{eq:gc.10}
	\mathrm{d}s^2 = f^{-1}\left[\left(\mathrm{d}\rho^2+\mathrm{d}\zeta^2\right)+\rho^2\mathrm{d}\varphi^2\right]-f\mathrm{d}t^2 \,,
\end{equation}
where
\begin{equation}
	%f^{1/2} = e^{U}=1-\epsilon^{-1}\alpha \,.
	f^{1/2} = e^{U}=1-\epsilon\alpha
\end{equation}
is the defining relation between the metric function $f$ and the electrostatic potential $\alpha$.
Due to the static spacetime the electromagnetic four-potential has only one non-vanishing component:
\begin{equation}
	A_a = (0,0,0,-\alpha) \,.
\end{equation}

Starting from line element (\ref{eq:gc.10}), the Einstein-Maxwell equations reduce to the surprisingly simple equation
\begin{equation} \label{eq:gc.6}
	\Delta e^{-U} = - 4\pi\mu^{\text{ECD}} e^{-3U} \,.
\end{equation}
By introducing a new potential $V$ and a redefined mass density $ \mu^{\text{ECD}}_{\text{st}}$ as
\begin{equation}
	V\coloneqq 1-e^{-U} \quad \text{and} \quad  \mu^{\text{ECD}}_{\text{st}}\coloneqq \mu^{\text{ECD}} e^{-3U} \,,
\end{equation}
equation (\ref{eq:gc.6}) can be transformed into the Poisson equation
\begin{equation}\label{eq:gc.8}
	\Delta V = 4\pi \mu^{\text{ECD}}_{\text{st}} \,.
\end{equation}
The solution can consequently be represented as a Poisson integral and based on its asymptotic behaviour we can identify the mass:
\begin{equation}\label{eq:gc.7}
	M =\int\mu^{\text{ECD}}_{\text{st}}\!\left(\rho, \zeta\right)\rho\,\mathrm{d}\rho\,\mathrm{d}\varphi\,\mathrm{d}\zeta \,.
\end{equation}

For ECD the mass density $\mu^{\text{ECD}}_{\text{st}}$ is not predetermined by the theory, but can rather be freely chosen. As a physically motivated toy model, we consider a specific ECD-disc configuration equipped with the mass density of a Maclaurin disc:
\begin{equation}\label{eq:gc.9}
	\mu^{\text{ECD}}_{\text{st}}=\mu^{\text{Mld}} \,.
\end{equation}
Note, that this means $\mu^{\text{ECD}}_{\text{st}}=e^{2U}\sigma^{\text{ECD}}_{\text{st}}(\rho)\delta(\zeta)=\sigma^{\text{Mld}}(\rho)\delta(\zeta)$. Choosing the mass density in this way is justified, since then the mass of the ECD-disc also coincides with the one of the Maclaurin disc: \mbox{$M =\int\mu^{\text{ECD}}\!\left(\rho, \zeta\right)\rho\,\mathrm{d}\rho\,\mathrm{d}\varphi\,\mathrm{d}\zeta = \int\mu^{\text{Mld}}_{\text{st}}\!\left(\rho, \zeta\right)\rho\,\mathrm{d}\rho\,\mathrm{d}\varphi\,\mathrm{d}\zeta$}.

In case of the Maclaurin disc, the equation of motion $\Delta U = 4\pi \mu^{\text{Mld}}$ leads to the solution $U\vert_{\nu=0} = \frac{1}{2}U_c\left(1+\eta^2\right)$, see equations (\ref{eq:gc.2}) and (\ref{eq:gc.4.5}). Analogously, for ECD the solution of (\ref{eq:gc.8}) with the chosen mass density (\ref{eq:gc.9}) is given by
\begin{equation}
	V\vert_{\nu=0} = \frac{1}{2}V_c\left(1+\eta^2\right) \,.
\end{equation}
However, now we get $V_c=-\frac{g^2}{1-g^2}$.

The proper spatial line element, stemming from (\ref{eq:gc.10}), for the chosen ECD-disc configuration in terms of elliptic coordinates then reads
\begin{equation}
	\mathrm{d}\sigma^2\Big\vert_{\nu=0} = f^{-1}\Big\vert_{\nu=0}\left(\rho_0^2\frac{\eta^2}{1-\eta^2}\mathrm{d}\eta^2 + \rho_0^2\left(1-\eta^2\right)\mathrm{d}\varphi^2\right) \,,
\end{equation}
where
%\begin{align}
%	&f^{-1/2} = 1-V \,,\notag \\
%	&V\vert_{\nu=0} = -\frac{1}{2}\frac{g^2}{1-g^2}\left(1+\eta^2\right) \,.
%\end{align}
\begin{equation*}
	f^{-1/2} = 1-V \,, \quad V\vert_{\nu=0} = -\frac{1}{2}\frac{g^2}{1-g^2}\left(1+\eta^2\right) \,.
\end{equation*}
It implies the following Gaussian curvature:
\begin{equation}
	K^{\text{ECD}} = \frac{1}{\rho_0^2}\frac{32\left(1-g^2\right)^2g^2}{\left[2-\left(1-\eta^2\right)g^2\right]^4} \,.
\end{equation}
\begin{figure}[htb]
	\centering
	\begin{tikzpicture}
		\node[anchor=south west, inner sep=0] (image) at (0,0) {\includegraphics[width=0.5\textwidth, trim={3.5cm 17.5cm 5.5cm 5.5cm}, clip]{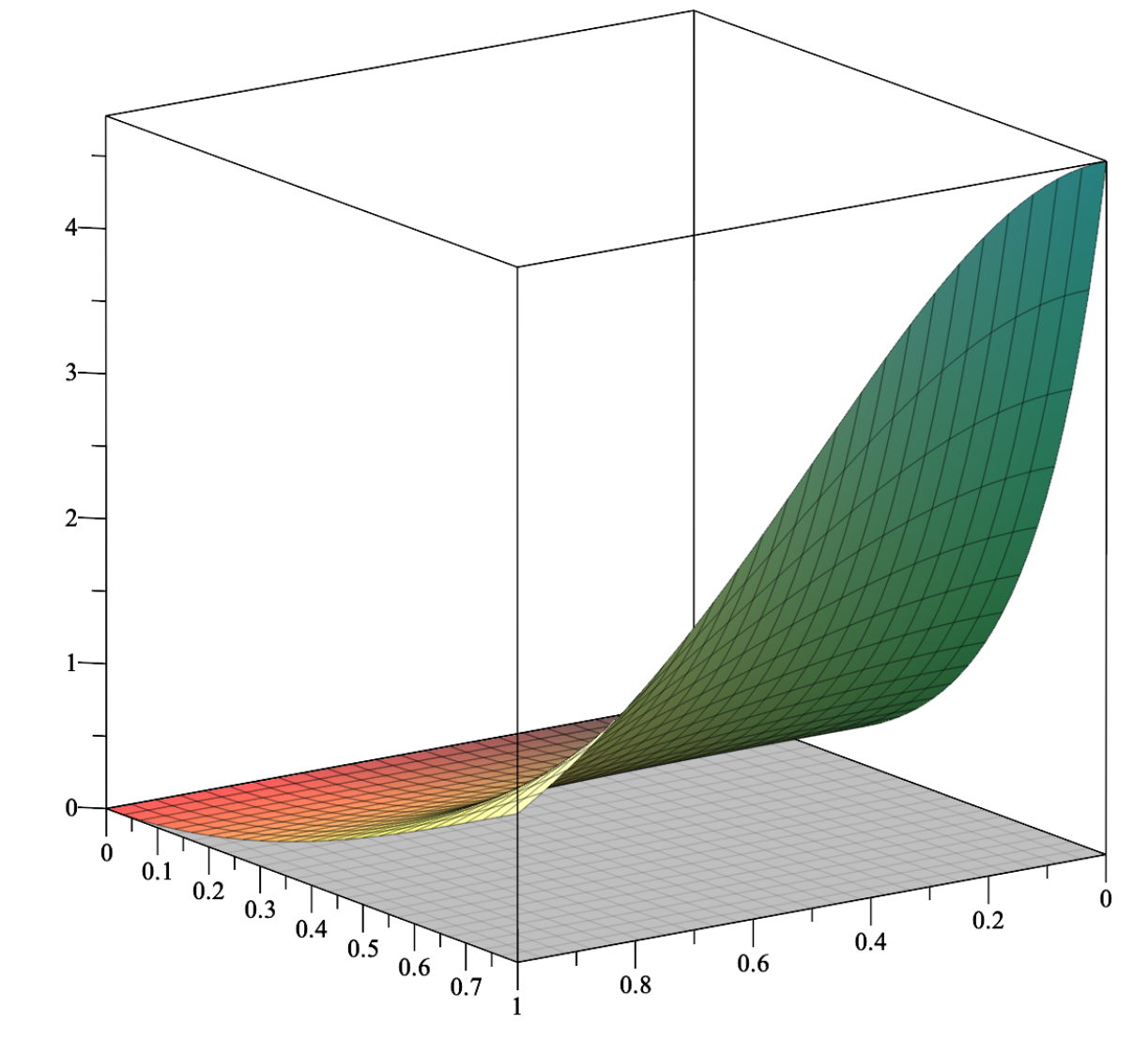}};
		\begin{scope}[x={(image.south east)}, y={(image.north west)}]
			\node[anchor=east] at (0.049, 0.567) {$\left(R_0^{\text{ECD}}\right)^2K^{\text{ECD}}$};
			\node at (0.273, 0.05) {$g$};
			\node at (0.713 ,0.05) {$\eta$};
		\end{scope}
	\end{tikzpicture}
	\caption{Normalized Gaussian Curvature of the chosen ECD-configuration.}
	\label{fig:6}
\end{figure}
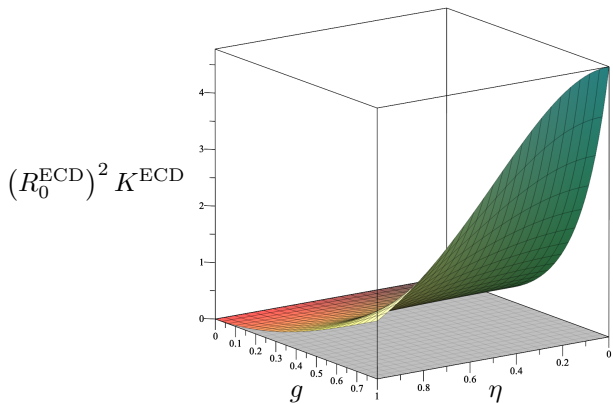
The corresponding plot of the normalized Gaussian curvature can be found in figure \ref{fig:6}.
There we introduced $R_0^{\text{ECD}}\coloneqq R^{\text{ECD}}(\eta=0)=\rho_0\frac{6-g^2}{6\left(1-g^2\right)}$.

Interestingly, in contrast to the charged rotating disc of dust (see \mbox{figure \ref{fig:4.2}}), the chosen ECD-disc configuration possesses higher (positive) curvature at the rim of the disc and lower at the centre. But nevertheless, it has a positive curvature in agreement with the $\epsilon\!=\!1$-limit of the charged rotating disc of dust.

To gain a better understanding of this radial curvature behaviour, we investigate and compare the proper surface mass densities of the so far discussed discs. The resulting proper surface mass densities are
\begin{align}
	\sigma_{\text{p}} &= \sigma \sqrt{fh^{-1}} = \frac{1}{2\pi}U'_{,\zeta}\Big\vert_{\zeta=0^+}\sqrt{fh^{-1}} \notag \\
	&= \frac{2}{\pi^2\rho_0} \left[ \eta g^{2} + \frac{1}{6}\left(\left(\frac{23}{3}\eta^{2}-5\right) \epsilon^{2}-7 \eta^{2}\right)  \eta g^{4}+\mathcal{O}\left(g^6\right) \right]\,, \\
	\sigma_{\text{p}}^{\text{Mld}} &= \sigma^{\text{Mld}} \notag \\
	&= \frac{2}{\pi^2\rho_0}g^2\eta \,, \\
	\sigma_{\text{p}}^{\text{ECD}} &= \sigma^{\text{ECD}}e^U \notag \\
	&= \frac{2}{\pi^2\rho_0}\left[1+\frac{1}{2}\frac{g^2}{1-g^2}\left(1+\eta^2\right)\right]^{-2}\frac{g^2}{1-g^2}\eta \notag \\
	&=\frac{2}{\pi^2\rho_0} \left[ \eta g^{2}-\eta^{3}g^{4}+\mathcal{O}\left(g^6\right) \right] \,.
\end{align}
As it should be, $\sigma_{\text{p}}^{\text{Mld}}$ represents the Newtonian limit of $\sigma_{\text{p}}$ and $\sigma_{\text{p}}^{\text{ECD}}$.

\begin{figure}[htb]
	\centering
	\begin{minipage}{0.45\textwidth}
		\centering
		\begin{tikzpicture}
			\node[anchor=south west, inner sep=0] (image) at (0,0) {\includegraphics[width=0.9\textwidth]{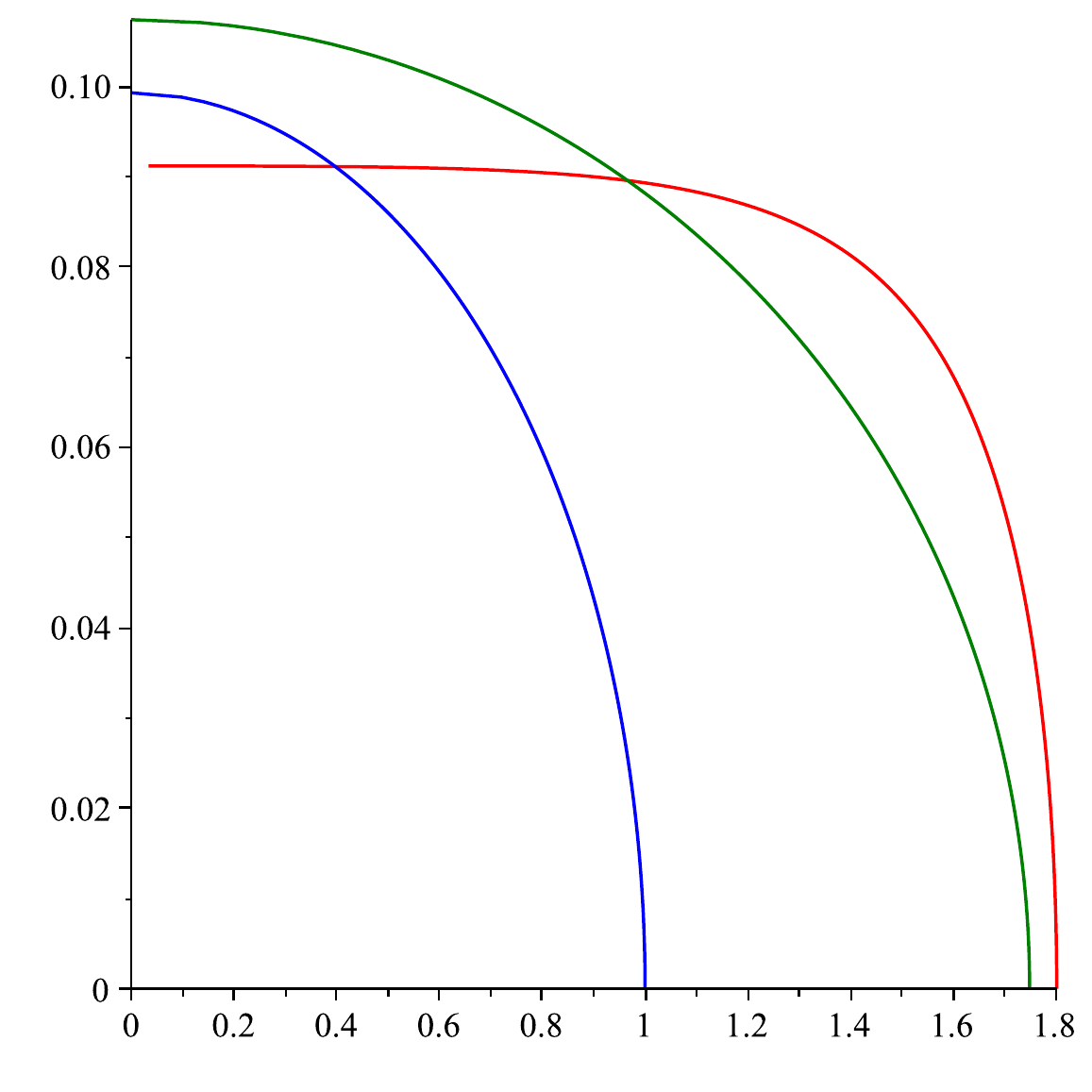}};
			\begin{scope}[x={(image.south east)}, y={(image.north west)}]
				\node[color=blue] at (0.33, 0.53) {$R_0^{\text{Mld}}\sigma_{\text{p}}^{\text{Mld}}$};
				\node[color=OliveGreen] at (0.61, 0.708) {$R_0\sigma_{\text{p}}$};
				\node[color=BrickRed] at (0.87, 0.854) {$R_0^{\text{ECD}}\sigma_{\text{p}}^{\text{ECD}}$};
				\node[color=blue] at (0.57, -0.02) {$R_0^{\text{Mld}}$};
				\node[color=OliveGreen] at (0.84, -0.02) {$R_{0}$};
				\node[color=BrickRed] at (0.99, -0.02) {$R_0^{\text{ECD}}$};
			\end{scope}
		\end{tikzpicture}
		\caption{Comparison of the normalized proper surface mass densities $R_0^{\text{Mld}}\sigma_{\text{p}}^{\text{Mld}}$, $R_{0}\sigma_{\text{p}}$ and $R_0^{\text{ECD}}\sigma_{\text{p}}^{\text{ECD}}$ for $g=0.7$ and $\epsilon=1$.}
		\label{fig:7.1}
	\end{minipage}\hfill
	\begin{minipage}{0.45\textwidth}
		\centering
		\begin{tikzpicture}
			\node[anchor=south west, inner sep=0] (image) at (0,0) {\includegraphics[width=0.9\textwidth]{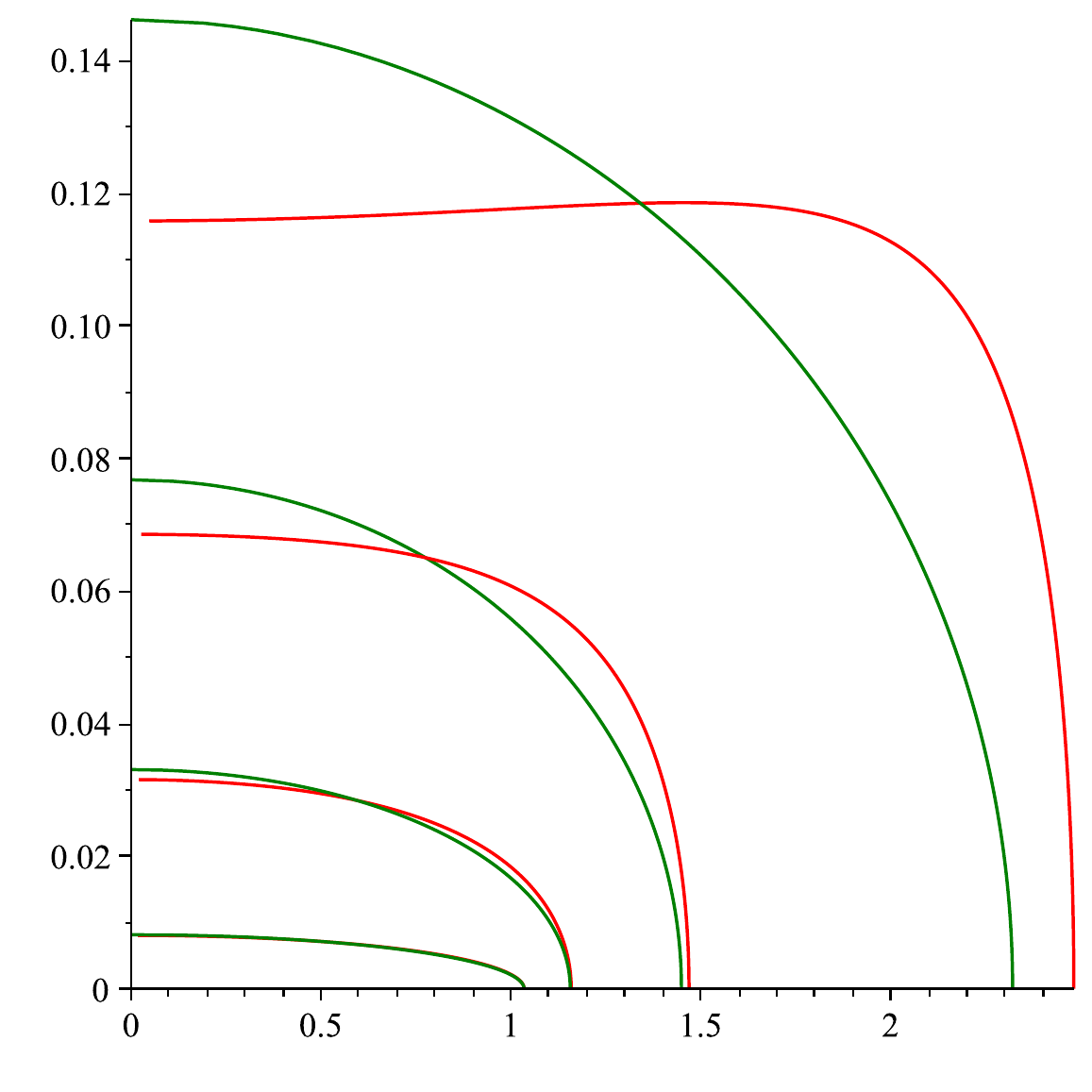}};
			\begin{scope}[x={(image.south east)}, y={(image.north west)}]
				\node[color=OliveGreen] at (0.6, 0.7) {$R_0\sigma_{\text{p}}$};
				\node[color=BrickRed] at (0.86, 0.866) {$R_0^{\text{ECD}}\sigma_{\text{p}}^{\text{ECD}}$};
				\node[color=OliveGreen] at (0.84, -0.02) {$R_{0}$};
				\node[color=BrickRed] at (0.99, -0.02) {$R_0^{\text{ECD}}$};
				\node at (0.25, 0.879) {\footnotesize $g=0.8$};
				\node at (0.25, 0.607) {\footnotesize $g=0.6$};
				\node at (0.25, 0.34) {\footnotesize $g=0.4$};
				\node at (0.25, 0.187) {\footnotesize $g=0.2$};
			\end{scope}
		\end{tikzpicture}
		\caption{Comparison of the normalized proper surface mass densities $R_{0}\sigma_{\text{p}}$ and $R_0^{\text{ECD}}\sigma_{\text{p}}^{\text{ECD}}$ for different values of $g$ and $\epsilon=1$.}
		\label{fig:7.2}
	\end{minipage}
\end{figure}

In figure \ref{fig:7.1} all three normalized proper surface mass densities are plotted and in \mbox{figure \ref{fig:7.2}} $R_{0}\sigma_{\text{p}}\vert_{\epsilon=1}$ and $R_0^{\text{ECD}}\sigma_{\text{p}}^{\text{ECD}}$ are depicted for different values of the relativity parameter $g$.

From figure \ref{fig:7.2} it is evident that $\sigma_{\text{p}}^{\text{ECD}}$ is denser at the rim and less dense at the centre compared to $\sigma_{\text{p}}\vert_{\epsilon=1}$. The higher $g$, the more extreme is this behaviour of the chosen ECD-configuration.
This dominance of the proper surface mass density $\sigma_{\text{p}}^{\text{ECD}}$ at the rim as opposed to the centre leads to the observed higher curvature $K^{\text{ECD}}$ at the rim. In contrast, the curvature $K\vert_{\epsilon=1}$ generated by the surface mass density $\sigma_{\text{p}}\vert_{\epsilon=1}$ of the maximally charged disc of dust increases towards the centre.

\subsection{Gaussian curvature of the uncharged disc of dust}\label{subsec5.4}

By means of the inverse scattering method that originates from soliton theory, the global problem of the uniformly rotating disc of dust without charge was solved rigorously by Neugebauer and Meinel \cite{Neugebauer:1993ct, PhysRevLett.75.3046, RFE}.

An essential part of this method is to formulate and to solve a Riemann-Hilbert problem. It turns out that this Riemann-Hilbert problem has a unique solution in the parameter region $0<\mu<\mu_0\coloneqq4.629...\,$, where $\mu\coloneqq 2\left(\rho_0\Omega\right)^2e^{-2V_0}$ and $V_0\coloneqq U'(\rho=0,\zeta=0)$. $\mu\to 0$ corresponds to the Newtonian limit and for $\mu\to\mu_0$ the formation of a Kerr-black hole was proven \cite{Neugebauer:1993ct, PhysRevLett.75.3046}, see also \cite{Meinel2002}.

Analogous to the charged disc, the line element can globally be written in Weyl-Lewis-Papapetrou form:
\begin{equation}
\mathrm{d}s^2 = e^{-2U'}\left[ e^{2k'}\left( \mathrm{d}\rho^2 + \mathrm{d}\zeta^2 \right) + \rho^2\mathrm{d}\varphi'^2 \right] - e^{2U'}\left( \mathrm{d}t + a\, \mathrm{d}\varphi' \right)^2 \,.
\end{equation}
Denoting $x\coloneqq\frac{\rho}{\rho_0}$, as used in the literature above, the resulting proper spatial line element reads
\begin{equation}
	\mathrm{d}\sigma^2\Big\vert_{\Sigma_2} = e^{-2\left(U-k\right)}\rho_{0}^2\,\mathrm{d}x^2 + e^{-2V_0}\rho_{0}^2 x^2\mathrm{d}\varphi'^2 \,,
\end{equation}
utilising the boundary condition $e^{2U'}=e^{2V_0}$ and the transformation law \mbox{$k'-U'=k-U$}.

%For the associated Gaussian curvature we obtain the compact formula
%\begin{equation}
%	K^{\text{uncharged}} = -\frac{1}{2\rho_{0}^2x}\frac{\partial}{\partial x}e^{2\left(U-k\right)} \,,
%\end{equation}
%where the analytic solution of $e^{-2\left(U-k\right)}=g_{\rho\rho}\left(\mu,x\right)$ is given in terms of theta functions as well as elliptic and complete elliptic integrals, see, e.g., \cite{RFE}.

\begin{figure}[htb]
	\centering
	\begin{minipage}{0.45\textwidth}
		\centering
		\begin{tikzpicture}
			\node[anchor=south west, inner sep=0] (image) at (0,0) {\includegraphics[width=1\textwidth, clip]{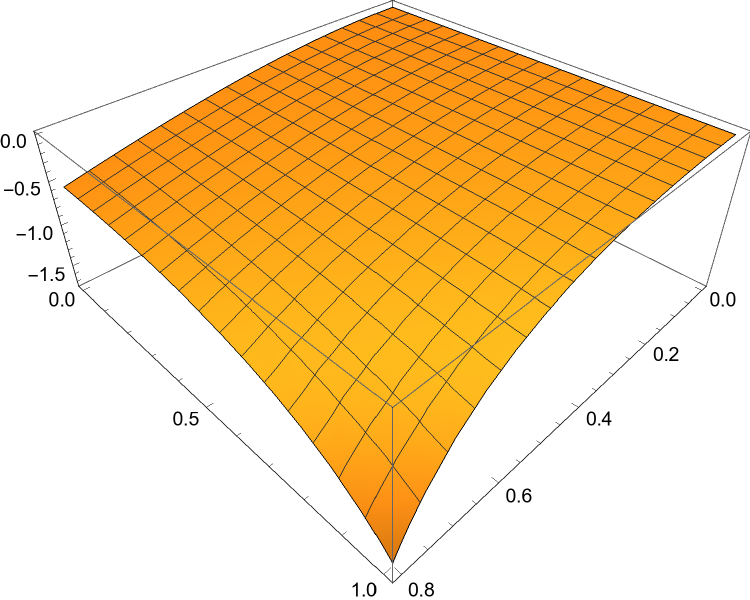}};
			\begin{scope}[x={(image.south east)}, y={(image.north west)}]
				\node [anchor=west] at (-0.02, 1.069) {$\left(R_{0}^{\text{uncharged}}\right)^2K^{\text{uncharged}}$};
				\node at (0.269, 0.215) {$x$};
				\node at (0.78 ,0.21) {$g$};
			\end{scope}
		\end{tikzpicture}
		\caption{Normalized Gaussian curvature of the uncharged disc of dust.}
		\label{fig:8.1}
	\end{minipage}\hfill
	\begin{minipage}{0.45\textwidth}
		\centering
		\begin{tikzpicture}
			\node[anchor=south west, inner sep=0] (image) at (0,0) {\includegraphics[width=1\textwidth, trim={3.5cm 17.5cm 5.5cm 5.5cm}, clip]{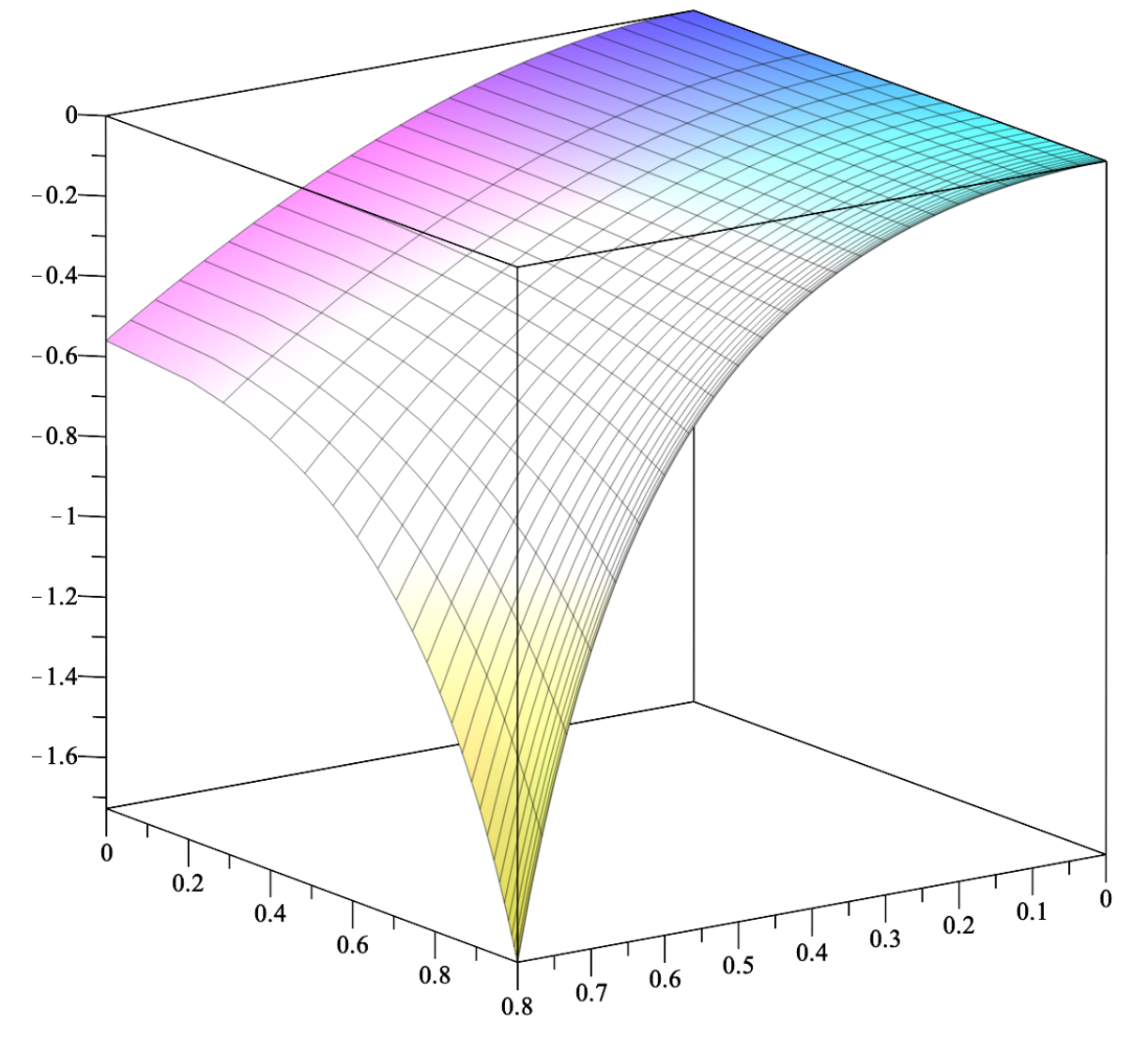}};
			\begin{scope}[x={(image.south east)}, y={(image.north west)}]
				\node[anchor=east] at (0.035, 0.567) {$R_{0}^{2}K$};
				\node at (0.273, 0.05) {$x$};
				\node at (0.713 ,0.034) {$g$};
			\end{scope}
		\end{tikzpicture}
		\caption{Normalized Gaussian curvature of the charged disc of dust in the limit of vanishing charge.}
		\label{fig:8.2}
	\end{minipage}
\end{figure}

For the associated Gaussian curvature we obtain the compact formula
\begin{equation}
	K^{\text{uncharged}} = - \frac{1}{2\rho_0^2 x}  \frac{\partial}{\partial x} e^{2(U-k)}
	%= - \frac{1}{2\rho_0^2 x}  \frac{\partial}{\partial x} e^{2(U'-k')}
	= - \frac{1}{2\rho_0^2 x}  \frac{\partial}{\partial x} e^{2(V_0-k')} \,.
\end{equation}
Hence the normalized Gaussian curvature plotted in figure \ref{fig:8.1} can be calculated by
\begin{equation}
	\left(R_{0}^{\text{uncharged}}\right)^{2}K^{\text{uncharged}}(g,x) = -2\mu k'_{,\tilde{\mu}}(\tilde{\mu})
	\frac{\left[\int_0^1 dx\, e^{k'}(\tilde{\mu}) \right]^2}{e^{2k'}(\tilde{\mu})} \,,
\end{equation}
with $\tilde{\mu}:=\mu(1-x^2)$ and the relation $g(\mu) = \sqrt{1-e^{V_0}(\mu)}$ between the
parameters $g$ and $\mu$. The only occurring metric potential $k'$
is related (within the disc) to the Ernst potential $e^{2V_0} + i b_0$ in the origin ($x=0$) by
\begin{equation}
	k'_{,\tilde{\mu}}(\tilde{\mu}) = \frac{1}{4}
	\frac{\left[{e^{2V_0}}_{,\tilde{\mu}}(\tilde{\mu} )\right]^2 +  \left[{b_0}_{,\tilde{\mu}}(\tilde{\mu})\right]^2}
	{e^{2V_0}(\tilde{\mu})} \,.
\end{equation}
These parameter functions $e^{2V_0}$ and $b_0$ itself are given in terms of Jacobi's elliptic functions.

Comparing the plot of the normalized Gaussian curvature of the uncharged disc (exact solution), figure \ref{fig:8.1}, with the one of the charged disc evaluated at $\epsilon=0$ (post-Newtonian expansion up to tenth order), figure \ref{fig:8.2}, shows a good qualitative agreement.

Checking the more conclusive numerical values, certifies an excellent coincidence between the (exact) analytic and the semi-analytic solution for $\epsilon=0$. %Apart from small deviations, the numerical values for $R_{0}^2K\vert_{\epsilon=0}$ and $\left(R_{0}^{\text{uncharged}}\right)^{2}K^{\text{uncharged}}$ match for $g=0.6$ up to and including the forth decimal place, for  $g=0.7$ up to and including the third and for  $g=0.8$ still up to and including the first, see appendix \ref{secB}.
Averaged over the disc (using $x=\{0, 0.3, 0.7, 1\}$), the percent deviation of $R_{0}^2K\vert_{\epsilon=0}$ from $\left(R_{0}^{\text{uncharged}}\right)^{2}K^{\text{uncharged}}$ is $\num{2.29e-6}$ for $g=0.6$, $\num{3.50e-5}$ for $g=0.7$ and still only $\num{4.72e-4}$ for $g=0.8$. See appendix \ref{secB} for the corresponding numerical values of the normalized Gaussian curvature.

\subsection{Visualization}\label{subsec5.5}

Similar to Flamm's paraboloid in case of Schwarzschild spacetime, we want to visualize the spatial curvature of the charged disc of dust by an isometric embedding. 

The isometric embedding of the proper 2-dimensional disc space, characterized by (\ref{eq:gc.11}), into 3-dimensional Euclidean space, furnished with the line element \mbox{$\mathrm{d}l^2=\mathrm{d}r^2+r^2\mathrm{d}\phi^2+\mathrm{d}z^2$} is achieved by the identifications
\begin{align}
	&\phi = \varphi' \,, \\
	&r = \frac{C'(R)}{2\pi} \,, \\
	&z(R) = \int_{0}^{R}\left[1-\frac{C'(\tilde{R})_{,\tilde{R}}}{2\pi} \right]^{1/2}\mathrm{d}\tilde{R} \,,
\end{align}
if the condition $C'(R)_{,R}\leq 2\pi$ is satisfied.

This means that this embedding only works for sufficiently high values of $\epsilon$.
Based on to the fact that the curvature becomes first negative at the centre of the disc by lowering $\epsilon$ (if initially it is positive everywhere), see section \ref{subsec5.6}, it can be shown that $K\geq 0$ is not only a sufficient but also a necessary condition for the embedding constraint, $C'(R)_{,R}\leq 2\pi$, to be fulfilled.

\begin{figure}[htb]
	\centering
	\begin{tikzpicture}
		\node[anchor=south west, inner sep=0] (image) at (0,0) {\includegraphics[width=0.7\textwidth, trim={3.5cm 22.5cm 5.5cm 10.35cm}, clip]{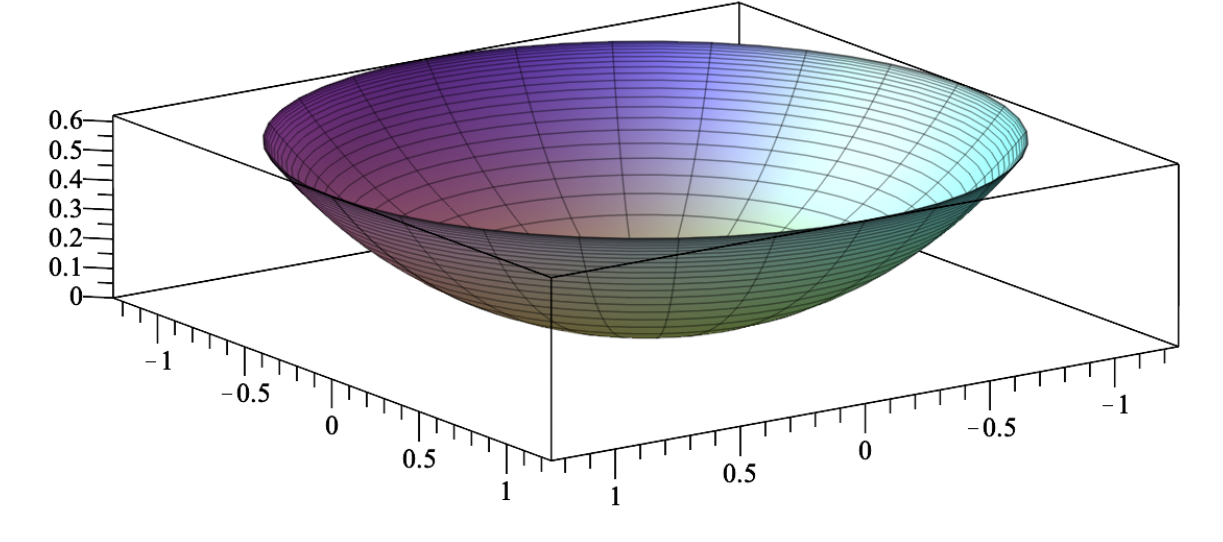}};
		\begin{scope}[x={(image.south east)}, y={(image.north west)}]
			\node[anchor=east] at (0.03, 0.62) {$z(r)$};
		\end{scope}
	\end{tikzpicture}
	\caption{Isometric embedding of the proper 2-dimensional disc space into 3-dimensional Euclidean space for $g=0.6$ and $\epsilon=1$. The scaling parameter $\rho_{0}$ is set to $1$.}
	\label{fig:9}
\end{figure}

In figure \ref{fig:9} the embedding of the charged disc of dust is depicted for the values $g=0.6$ and $\epsilon=1$.

\subsection{Transition curves}\label{subsec5.6}

\begin{figure}[htb]
	\centering
	\begin{minipage}{0.45\textwidth}
		\centering
		\begin{tikzpicture}
			\node[anchor=south west, inner sep=0] (image) at (0,0) {\includegraphics[width=1\textwidth, trim={5.5cm 18.5cm 6.5cm 7.5cm}, clip]{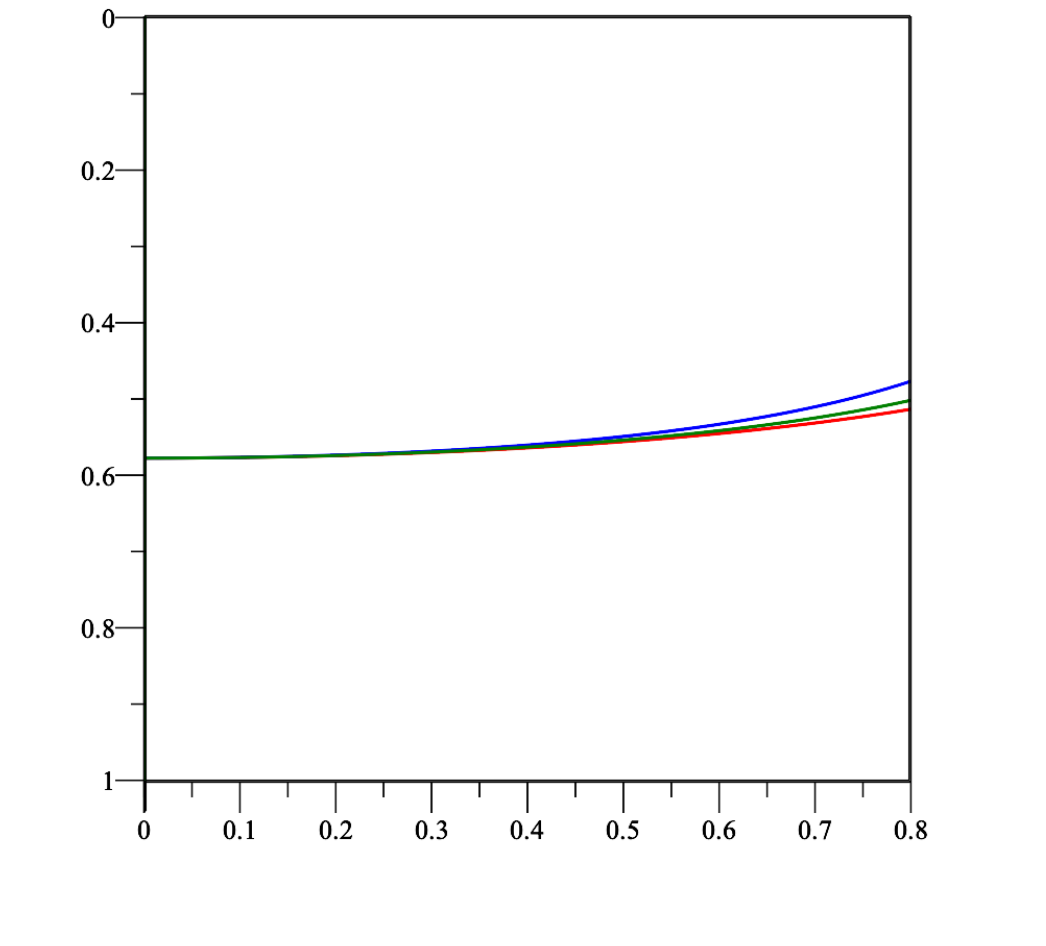}};
			\begin{scope}[x={(image.south east)}, y={(image.north west)}]
				\node [anchor=east] at (0.06, 0.577) {$\epsilon$};
				\node at (0.5, 0.04) {$g$};
				\node[color=blue] at (0.91, 0.64) {\footnotesize 2)};
				\node[color=OliveGreen] at (0.91, 0.58) {\footnotesize 3)};
				\node[color=BrickRed] at (0.91, 0.52) {\footnotesize 1)};
			\end{scope}
		\end{tikzpicture}
		\caption{Conditions 1),  2), and 3) lead to very similar transition curves in the parameter space, here depicted for $\eta=0$.}
		\label{fig:10.1}
	\end{minipage}\hfill
	\begin{minipage}{0.45\textwidth}
		\centering
		\begin{tikzpicture}
			\node[anchor=south west, inner sep=0] (image) at (0,0) {\includegraphics[width=1\textwidth, trim={5.5cm 18.5cm 6.5cm 7.5cm}, clip]{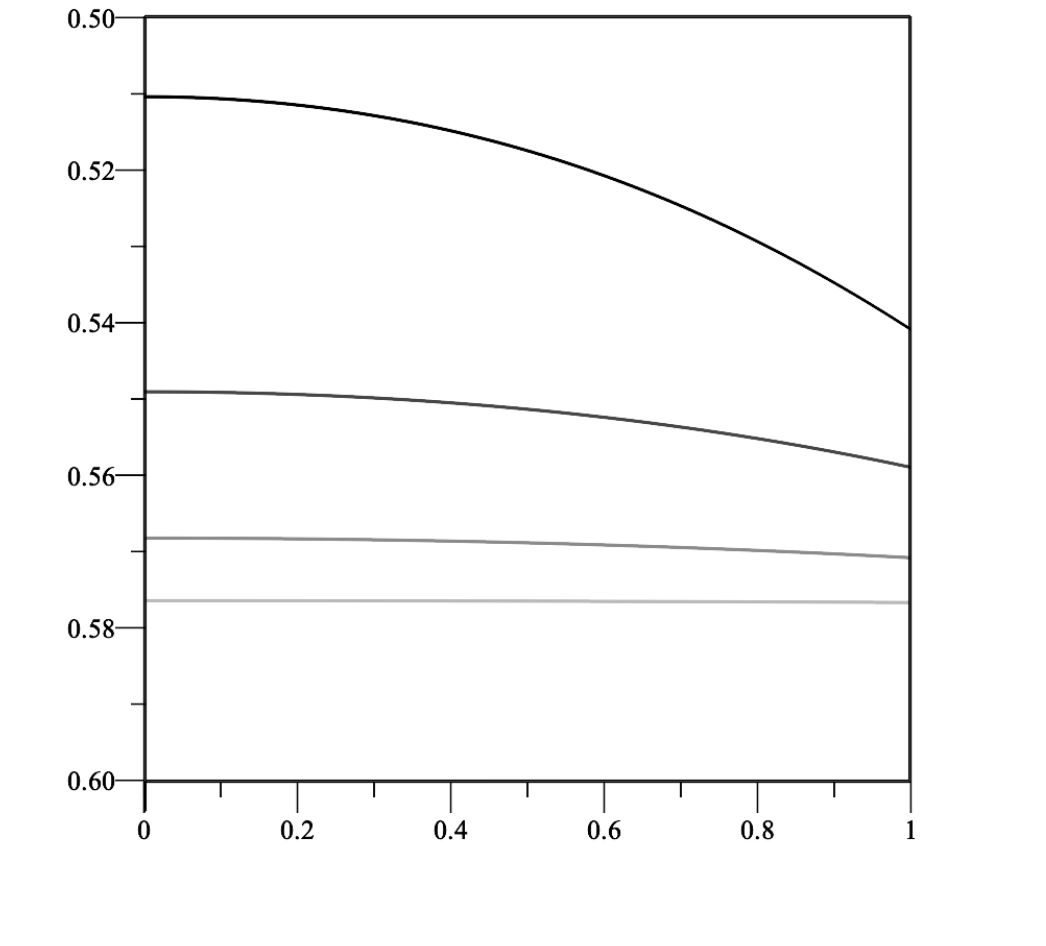}};
			\begin{scope}[x={(image.south east)}, y={(image.north west)}]
				\node [anchor=east] at (0.06, 0.577) {$\epsilon$};
				\node at (0.5, 0.04) {$\eta$};
				\node[color=black] at (0.97, 0.653) {\footnotesize $g=0.7$};
				\node[color=black!83] at (0.97, 0.509) {\footnotesize $g=0.5$};
				\node[color=black!67] at (0.97, 0.42) {\footnotesize $g=0.3$};
				\node[color=black!50] at (0.97, 0.36) {\footnotesize $g=0.1$};
			\end{scope}
		\end{tikzpicture}
		\caption{Radial dependence of the curve $K=0$ plotted for different values of the relativistic parameter $g$.}
		\label{fig:10.2}
	\end{minipage}
\end{figure}

In three different cases we have seen the occurrence of a transition curve in the parameter space $(\epsilon,g)$:
%\begin{enumerate}[label=\arabic*)]
%	\item $\frac{C'}{R'}=2\pi$ \,,
%	\item $K=0$ \,, 
%	\item $C'_{,R}=2\pi$ \,.
%\end{enumerate}
\begin{equation*}
	\text{1)}\;\,\frac{C'}{R'}=2\pi \;, \qquad
	\text{2)}\;\,K=0 \;, \qquad
	\text{3)}\;\,C'_{,R}=2\pi \;.
\end{equation*}
%It turns out that all those three conditions are equivalent and indeed all the curves are identical. This is also backed up by figure \ref{fig:10.1}.
It turns out that these transition curves look very similar, almost identical, as shown in figure \ref{fig:10.1}.

The radial dependence of the transition curve $K=0$ (the others being very similar) can be read off from \mbox{figure \ref{fig:10.2}}. As can be seen there, changing $\epsilon$ does not cause a transition throughout the whole disc at the same time. In fact, by starting with positive curvature, i.e.\ high values of $\epsilon$, the transition to negative curvature happens first at the centre and last at the rim.

%\section{Discussion}\label{sec12}

%Discussions should be brief and focused. In some disciplines use of Discussion or `Conclusion' is interchangeable. It is not mandatory to use both. Some journals prefer a section `Results and Discussion' followed by a section `Conclusion'. Please refer to Journal-level guidance for any specific requirements. 

\section{Discussion}\label{sec6}

A central assumption in cosmology is that the universe is spatially homogeneous and isotropic on large length scales. This is the cosmological principle and indeed observations confirm that it is fulfilled on scales of about \mbox{100 \si{Mpc}}.
%In cosmology the cosmological principle tells us that the universe is homogeneous and isotropic on large scales. Indeed, observations confirm this for scales of about 100 \si{Mpc}. 
Using the cosmological principle one can straightforwardly derive the Friedmann-Robertson-Walker metric that reduces the Einstein equations to the famous Friedmann equations. Due to homogeneity and isotropy the solutions exhibit a globally constant geometry with only three possible spatial curvatures: flat ($k=0$), positive ($k=+1$) or negative ($k=-1$). Remarkably, observations of the CMB suggest that our universe might indeed be flat. The first Friedmann equation reveals that there is a critical energy density that leads to a flat universe. If the density is higher than the critical density the universe is positively curved and if it is lower it is negatively curved. For more details see, e.g., \cite{mukhanov2005}.

In some sense the proper spatial geometry of the charged rotating disc of dust behaves analogous to the spatial curvature of the Friedmann universe. High values of the specific charge $\epsilon$ (defined as the ratio of electric charge to baryonic mass) cause positive and low values negative curvature. Furthermore, for given $g$ and $\eta$ there is a critical value of $\epsilon$ that gives rise to Euclidean geometry. However, unlike in Friedmann cosmology the geometry of the charged disc of dust is not globally constant. The geometry remains unchanged only in angular direction, but not in radial one. A beautiful exception of this represents the Newtonian limit with a radially independent curvature. In fact, for the critical value $\epsilon=\frac{1}{\sqrt{3}}$ the disc is globally flat.

\bmhead{Acknowledgments}

This work has been funded by the Deutsche
Forschungsgemeinschaft (DFG) under Grant No.
406116891 within the Research Training Group RTG
2522/1. 
The authors would like to thank \mbox{Martin} Breithaupt for the provided results of the post-Newtonian expansion up to tenth order.

\section*{Declarations}

Conflict of interest: The authors have no relevant financial or non-financial interests to disclose. \\
Data availability: The datasets generated during and/or analysed during the current study are available from the corresponding author on reasonable request.

\begin{appendices}

\section{Expansions up to fifth order}\label{secA}

\begin{align*}
	\frac{1}{2\pi}&\frac{C'}{R'}\left(\eta=0\right)  \\
	={}&1-\frac{1}{2}\left(\epsilon^{2}-\frac{1}{3}\right) g^{2} + \frac{5}{24 \pi^{2}} \left[\left(\epsilon^{4}-\frac{52}{75} \epsilon^{2}-\frac{53}{75}\right) \pi^{2}-\frac{224 \epsilon^{2}}{25}+\frac{224}{25}\right] g^{4} \\
	&-\frac{223}{1920 \pi^{2}} \left[\left(\epsilon^{6}-\frac{3457}{892} \epsilon^{4}+\frac{113453}{112392} \epsilon^{2}+\frac{31336}{14049}\right) \pi^{2} \right. \\
	&\left. -\,\frac{2560}{669} \left(\epsilon ^2-1\right) \left(\epsilon^{4}-\frac{313}{70} \epsilon^{2}-\frac{257}{42}\right)\right] g^{6} \\
	&+\frac{3145}{48384 \pi^{4}} \left[\left(\epsilon^{8}-\frac{313007839}{40256000} \epsilon^{6}+\frac{3055678201}{193228800} \epsilon^{4}-\frac{6005877459}{644096000} \epsilon^{2}-\frac{29078}{235875}\right) \pi^{4} \right. \\
	&\left.-\,\frac{111104}{15725} \left(\epsilon^{2} -1\right)  \left(\epsilon^{6}-\frac{69012347}{9374400} \epsilon^{4}+\frac{113190529}{24998400} \epsilon^{2}+\frac{78691}{13020}\right) \pi^{2} \right. \\
	&\left.+\,\frac{802816}{15725} \left(\epsilon^ 2-1\right)^{2} \left(\epsilon^{2}-\frac{15227}{1960}\right)\right] g^{8} \\ 
	&-\frac{2815391}{66355200 \pi^{4}} \left[\left(\epsilon^{10}-\frac{379034234759}{41622740544} \epsilon^{8}+\frac{95728571097211}{3874698756096} \epsilon^{6} \right.\right. \\
	&\left.\left.-\,\frac{103871864922476873}{2727787924291584} \epsilon^{4}+\frac{14955289324645259}{495961440780288} \epsilon^{2}-\frac{5355807104}{650355321}\right) \pi^{4} \right. \\
	&\left.-\,\frac{238424064}{19707737} \left(\epsilon^ 2-1\right) \left(\epsilon^{8}-\frac{10000383870041}{1084307938560} \epsilon^{6}+\frac{104207736848353}{5046960586752} \epsilon^{4} \right.\right. \\
	&\left.\left.-\,\frac{49539410937953809}{5921767088455680} \epsilon^{2}-\frac{91863454}{13446279}\right) \pi^{2} \right. \\
	&\left.+\,\frac{78643200}{2815391} \left(\epsilon^2-1\right)^{2}  \left(\epsilon^{6}-\frac{4169}{350} \epsilon^{4}+\frac{6055401553}{271656000} \epsilon^{2}+\frac{2656693}{46200}\right)\right] g^{10}
\end{align*}
\begin{align*}
	R_{0}^2&K\left(\eta=0\right)  \\
	={}&\left(3 \epsilon^{2}-1\right) g^{2}-5 \left(\epsilon^{2}-\frac{7}{30}\right) \left(\epsilon^2-1\right) g^{4} \\
	&+\,\frac{1703}{320 \pi^{2}} \left[\left(\epsilon^{6}-\frac{52261}{20436} \epsilon^{4}+\frac{219665}{122616} \epsilon^{2}-\frac{3968}{15327}\right) \pi^{2}-\frac{2560 \epsilon^{6}}{5109}+\frac{34816 \epsilon^{4}}{15327} \right. \\
	&\left.-\,\frac{11776 \epsilon^{2}}{5109}+\frac{8192}{15327}\right] g^{6} \\
	&-\,\frac{9769}{2016 \pi^{2}} \left[\left(\epsilon^{8}-\frac{274414949}{60020736} \epsilon^{6}+\frac{119546257}{18756480} \epsilon^{4}-\frac{249550409}{80027648} \epsilon^{2}+\frac{16824}{48845}\right) \pi^{2} \right. \\
	&\left.-\,\frac{94976}{48845} \left(\epsilon^{6}-\frac{964087}{142464} \epsilon^{4}+\frac{1525885}{284928} \epsilon^{2}-\frac{856}{1113}\right) \left(\epsilon^2 -1\right)\right] g^{8} \\
	&+\,\frac{316977977}{77414400 \pi^{4}} \left[\left(-\frac{158969856}{316977977}+\epsilon^{10}-\frac{425179610969}{60859771584} \epsilon^{8}+\frac{1000003956304379}{62320406102016} \epsilon^{6} \right.\right. \\
	&\left.\left.-\,\frac{61149890345347447}{3988505990529024} \epsilon^{4}+\frac{15274226506312301}{2659003993686016} \epsilon^{2}\right) \pi^{4}
	-\frac{1296420864}{316977977} \left(\epsilon^2 -1\right) \left(\epsilon^{8}    \vphantom{\frac{1129856}{949527}} \right.\right. \\
	&\left.\left.-\,\frac{32218659817}{3646183680} \epsilon^{6}+\frac{47760895613989}{2800269066240} \epsilon^{4}-\frac{566746281193567}{59739073413120} \epsilon^{2}+\frac{1129856}{949527}\right) \pi^{2} \right. \\
	&\left.+\,\frac{550502400}{316977977} \left(\epsilon^{4}-\frac{56}{5} \epsilon^{2}+\frac{5296}{175}\right) \left(\epsilon^2-1\right)^{2} \left(\epsilon^{2}-\frac{1}{3}\right)\right] g^{10}
\end{align*}
%\newpage%\FloatBarrier

\section{Gaussian curvature}\label{secB}

\begin{table}[!h]
	\centering
		\begin{minipage}{7.87cm}
			\caption{Gaussian curvature: uncharged disc (exact solution) versus charged disc in the limit $\epsilon\to0$ (series expansion up to tenth order).}\label{tab1}
			\begin{tabular*}{\textwidth}{@{\extracolsep{\fill}}lcccccc@{\extracolsep{\fill}}}
				\toprule
				&		&	$K^{\text{uncharged}}\left(R_{0}^{\text{uncharged}}\right)^2$	&	$KR_{0}^{2}\Big\vert_{\epsilon\to0}$ \\
				\midrule
				$g=0.6$	& 	$x=0$	&	- 0.333602	&	- 0.333602   \\		%-0.3336020963220731
							& 	$x=0.3$	&	- 0.349614	&	 - 0.349614	\\		%-0.3496141992110335
							& 	$x=0.7$	&	- 0.434073  & 	- 0.434072  \\		%-0.4340723982857976
							& 	$x=1$	&	- 0.584301  & 	- 0.584297  \\		%-0.5842972371947541
				\midrule
				$g=0.7$	& 	$x=0$	&	- 0.441980	&	- 0.441979  \\		%-0.4419793689296314
				& 	$x=0.3$	&	- 0.471784	& 	- 0.471784  \\		%-0.4717839561419511
				& 	$x=0.7$	&	- 0.640493  & 	- 0.640487  \\		%-0.6404872042732119
				& 	$x=1$	&	- 0.988126  & 	- 0.987999	\\		%-0.9879994909137558
				\midrule
				$g=0.8$	& 	$x=0$	&	- 0.560113	&	- 0.560108  \\		%-0.5601075062911154
				& 	$x=0.3$	&	- 0.611598	& 	- 0.611586  \\		%-0.6115858712884469
				& 	$x=0.7$	&	- 0.930419  & 	- 0.930281  \\		%-0.930280973723539
				& 	$x=1$	&	- 1.73047  &	- 1.72751	 \\		%-1.727508717154620
				\botrule
			\end{tabular*}
		\end{minipage}
\end{table}
\FloatBarrier

\end{appendices}


\bibliography{geometry-bibliography}% common bib file


\end{document}